\documentclass[12pt,a4paper,fleqn]{article} 
\usepackage{exscale,amssymb}
\usepackage[fleqn]{amsmath}
\usepackage[english]{babel}
\usepackage{rawfonts,latexsym,lscape,epsfig,multirow,lscape,graphics,bm, supertabular}
\usepackage{dcolumn}
\usepackage{pdfcomment}
\usepackage{tabularx}
\usepackage{eurosym}
\usepackage{arydshln}
\usepackage{xspace}
\newcolumntype{L}[1]{>{\raggedright\arraybackslash}p{#1}} 
\newcolumntype{C}[1]{>{\centering\arraybackslash}p{#1}} 
\newcolumntype{R}[1]{>{\raggedleft\arraybackslash}p{#1}} 

\newfont{\tabfont}{cmr7 at 7pt}
\everymath{\displaystyle}
\newcolumntype{L}{>{$}l<{$}}
\newcolumntype{C}{>{$}c<{$}}

 \newcommand*{\ov}[1]{	$\m@th\overline{\mbox{#1}}$}
 \newcommand*{\ovA}[1]{$\m@th\overline{\mbox{#1}\raisebox{3mm}{}}$}
\newcommand{\ts}{\textsuperscript}

\usepackage{tabularx}
\usepackage{tocbibind} 
\usepackage{caption}
\usepackage{setspace}
\usepackage{amsmath}
\usepackage{amsfonts}
\usepackage{amssymb}
\usepackage{faktor}
\usepackage{polynom}
\usepackage{graphicx}
\usepackage{calc}
\usepackage{hyperref}
\usepackage{mathdots}
\usepackage{multirow}
\usepackage{bigdelim}
\usepackage{tikz}
\usepackage{color}
\usepackage{rotating,tabularx}
\usetikzlibrary{patterns}
\usetikzlibrary{arrows}
\usetikzlibrary{fadings}
\usetikzlibrary{patterns}
\usepackage{mathtools,bm}
\usepackage{setspace}
\usepackage[round, sort]{natbib}

\DeclareMathOperator*{\argmin}{arg\,min}

\DeclarePairedDelimiter\floor{\lfloor}{\rfloor}

\begin{document}

\title{Identifying Latent Structures in Maternal Employment: Evidence on the German Parental Benefit Reform}

\author{Sophie-Charlotte Klose\thanks{Department of Management, University of Duisburg-Essen, Germany,
		{\sl e-mail:} sophie-charlotte.klose@uni-due.de. I am grateful to my supervisor Prof. Dr. Marie Paul for the support and encouragement. Her numerous comments and suggestions have greatly improved the paper. I thank Prof. Dr. Bernd Fitzenberger, Prof. Dr. Marie Paul, and Arnim Seidlitz for the fruitful collaboration in the data preparation. Moreover, I gratefully acknowledge financial support by the German Science Foundation (DFG) through the
	priority program 1764 ``The German Labor Market in a Globalized World''.}}

\date{\textbf{\today}}

\thispagestyle{empty} \maketitle

\begin{abstract}
	\noindent This paper identifies latent group structures in the effect of motherhood on employment by employing the C-Lasso, a recently developed, purely data-driven classification method. Moreover, I assess how the introduction of the generous German parental benefit reform in 2007 affects the different cluster groups by taking advantage of an identification strategy that combines the sharp regression discontinuity design and hypothesis testing of predicted employment probabilities. The C-Lasso approach enables heterogeneous employment effects across mothers, which are classified into an a priori unknown number of cluster groups, each with its own group-specific effect. Using novel German administrative data, the C-Lasso identifies three different cluster groups pre- and post-reform. My findings reveal marked unobserved heterogeneity in maternal employment and that the reform affects the identified cluster groups' employment patterns differently.
\end{abstract}
\vspace{0,5cm}

\newpage
\doublespacing

\section{Introduction}
As women's labor market participation has increased over the past several decades while fertility rates have simultaneously decreased, policymakers face the challenge of encouraging fertility, and trying to keep women after childbirth close to the labor market (\cite{R19}). In almost all European economies, paid parental leave policies try to mitigate this challenge by providing employment protection and some extent of earnings replacement, which reduces child-rearing opportunity costs and facilitates the return to employment after childbearing.\footnote{For an overview of parental leaves in Europe see, e.g., \cite{pronzato2009return}.} The efficient design of these policies hinges on an accurate estimate of child age's effect on maternal labor supply.

Providing an accurate estimate of maternal employment is sophisticated because mothers' labor supply reactions following a child's birth are very heterogeneous. However, researchers usually disregard the heterogeneity and only estimate the average effect of child age on employment. Sometimes effect heterogeneity regarding observable characteristics like education or age is investigated (e.g., \cite{Angrist1998}; \cite{Fit13}). Some studies show that the impact of child age on maternal labor supply differs between high-earning (highly educated) and less-earning (low-educated) women (e.g., \cite{Troske2010a}; \cite{Troske2013}). Due to higher opportunity costs of childbearing, highly educated and high-earning women are typically more labor market responsive than their less-educated and lower-earning peers (\cite{Lund18}). This pattern shows up at the so-called ``baby gap'' (\cite{R19}): the negative relationship between education and completed fertility, which leads to a fertility differential between highly educated and less-educated women. 

Nevertheless, the employment effects may also differ regarding unobservable characteristics: having a child may affect different groups of women based on the data's latent structures very differently. Estimating average effects may hide important and potentially policy-relevant differences in the employment trajectories. Reforms may also change the composition of groups and alter how a child influences a mother's employment. 

In this paper, I focus specifically on identifying latent structures in the effect of motherhood on employment through data-determined grouping. This is an important issue because policymakers require the most accurate information on this effect to react appropriately. A machine learning approach is appealing for examining latent heterogeneity in maternal employment decisions because it does not require the specification of any modeling mechanism for the latent group structure. 

My goal in this paper is twofold. First, I examine how the heterogeneous patterns of mothers' return behavior to employment up to six years after first childbirth look like. For this purpose, I classify mothers into a priori number of unknown cluster-groups. The group membership of each mother is also a priori unknown. Within each group, the effects of child age on employment are the same, whereas, across the groups, the effects of child age on employment differ. The econometric approach I use is a recently developed, purely data-driven classification method called C-Lasso proposed by \cite{Su16}. It isolates three different clusters of mothers with similar effects of child age on employment: a group of mothers who return to employment quickly after first childbirth, a cluster of mothers who are on average less employed, and a cluster of mothers who take an extended family break after birth before returning to employment. This finding reveals pronounced unobserved heterogeneity in the impact of child age on employment across mothers. In contrast to the recently used, purely descriptive classification analysis of maternal career trajectories after first birth conducted by \cite{Fr16}, the C-Lasso technique achieves simultaneous group classification and oracle-efficient estimation of the coefficients in each group. Moreover, the C-Lasso method allows for the heterogeneity in the slope coefficients when the extent of the heterogeneity is unknown. To the best of my knowledge, the C-Lasso method has been used only once in the context of estimating heterogeneous effects of the minimum wage on employment within a hierarchical model (see \cite{Wange2019}). 

Second, I want to find out how the most recent German parental leave reform, as one of the most significant reforms in German family policies in the last decades, has affected mothers' heterogeneous employment patterns after the first birth. This reform substantially changed parental compensation for time absent from employment following childbirth. Implemented in 2007, the new parental benefit offers a generous earnings-dependent income replacement up to 14 months after childbirth or a 3,600 EUR basic cash transfer for women not in the labor force in the pre-birth period. In contrast, the benefit eligibility under the old scheme was means-tested and only targeted to low-income families. Nonetheless, although the new benefits scheme raised the financial incentives by up to 21,000 EUR, there are huge differences in the benefit transfers across mothers from different education and earning groups.

To measure the extent to which the different cluster-groups have reacted to this reform and offer a new look to the empirical evidence of policy impacts on maternal employment, I apply a sharp regression discontinuity (RD) design in the spirit of \cite{Kluv18}. To identify the reform effects within the C-Lasso methodology's panel structure model, I use an identification strategy that exploits the hypothesis testing of predicted employment probabilities. As my data set, I use a novel custom-shaped administrative data set for which precise information on childbirths from the German Pension Registry have been matched to female employment biographies in Germany.

If the parental leave reform generates heterogeneous reactions to how mothers' labor supply responds to the child's age, I should observe different changes in predicted employment probabilities within the three cluster-groups. My empirical analysis does uncover marked latent heterogeneity in the reform reactions across groups: The reform generates strong negative and long-standing effects for the medium-sized group, whereas it slightly encourages the largest group to take up employment immediately after benefit exhaustion.

My findings contribute to a developing body of dynamic structural life cycle models and quasi-experimental literature on labor supply, fertility, and family policies. Most of the previous papers in the dynamic structural life cycle literature (e.g., \cite{Car01}; \cite{Hys99}; \cite{Mich11}; \cite{Moffitt1984}) analyze female labor supply and fertility jointly and find that labor supply decisions are characterized by significant state dependence and unobserved heterogeneity but do not directly link their evidence to parental leave policies. The literature using dynamic life-cycle models traditionally specifies labor supply decisions as a threshold-crossing model and estimates a random-effects version of this model. In contrast, I apply the C-Lasso to a dynamic fixed effects probit model because this avoids the distribution of unobserved heterogeneity from being arbitrarily restricted and the correlation with the covariates, and it avoids the initial conditions problem. More recent papers (e.g., \cite{Troske2010a}; \cite{Troske2013}) estimate the causal effects of birth on subsequent employment of mothers by extending the aforementioned dynamic structural life-cycle model towards a framework that takes into account endogeneity of labor market and fertility decisions, heterogeneity of the effects of children and their correlation with fertility decisions, and correlation of sequential labor market decisions. Both papers concentrate on the heterogeneity in the timing and spacing of births around the average career path. Whereas the major part of the literature has assessed the effect heterogeneity of children on labor supply within a tight framing, \cite{Fr16} provide to my knowledge the only structural model evidence for more general latent heterogeneity in the labor market outcomes of mothers by modeling separate cluster-group-specific career trajectories. Using Bayesian Markov chain clustering analysis, the authors classify women into five cluster groups with very different long-run trajectories. Their approach is purely descriptive because the authors focus on clustering transition processes (in and out of parental leave, non-employment, and different forms of employment) rather than credible isolate causal pathways. Compared to this study, I can provide simultaneous group classification and estimation of the effect of child age on employment. Moreover, my identification strategy that combines the sharp RD design and hypothesis testing of predicted employment probabilities can isolate causal pathways of the most recent parental benefit reform -- albeit under strong assumptions that need to be imposed. 

The quasi-experimental literature on paid parental leave in Germany focuses largely on policy effects on maternal employment (see particularly \cite{Ber2011}; \cite{Greyer2015}; \cite{Kluve2013}; \cite{Kluv18}; \cite{Schoenberg2014}). All papers adopt a sharp RD design or a differences-in-differences design by exploiting a natural experiment to compare mothers who gave birth to children shortly before the parental leave reform (control group) with mothers who gave birth to children shortly after the reform (treatment group). The mothers from the control and treatment groups face different paid maternity leaves, both in terms of leave duration and transfers. Overall, there is a large consensus across the studies: The generous paid parental leave reform in 2007 generates a strong disincentive-to-work effect in the short run. \cite{Schoenberg2014} analyze the impact of five expansions in maternity leave coverage on mother's labor market outcomes after childbirth and find that each expansion reduces mothers' postbirth employment rates in the short-run, whereas the long-run effects are small. \cite{Greyer2015} find that especially mothers with higher pre-birth earnings after the reform have more incentives to stay at home in the first year after giving birth than before the reform. However, they find stronger incentives to work after benefit exhaustion for low-income mothers. \cite{Kluv18} also find beneficial medium-run employment effects of the parental benefit reform. The authors find considerable positive and statistically significant effects on maternal employment, ranging up to 10\%. The positive impacts are centered on mothers from the medium and high terciles of earning distribution, whereas low-income mothers do not benefit.

Most recent literature also assesses the impact on fertility (see \cite{R19}). Using the large differential changes in maternity leave benefits of the most recent parental leave reform in 2007 across education and income groups in a differences-in-differences design, \cite{R19} finds a positive and statistically significant impact of increased benefits on fertility, driven mainly by women at the middle and upper end of the earnings distribution.

I add to the quasi-experimental literature on parental leave policies along several important dimensions. Most importantly, I provide evidence on how the most recent German parental benefit reform affects the different isolated cluster-groups' employment patterns. These group-specific policy impacts on maternal employment are potentially more policy-relevant than estimating only the average policy impact on employment. Another limitation of the previous literature is that it mostly concentrates on effect heterogeneity of the policy change based on observables. In contrast to the aforementioned literature, on the one hand, I cannot decipher the main determinants of heterogeneity as it would be feasible in an analysis based on observables. However, on the other hand, my approach provides a far more complete picture of how policy change affects maternal employment by taking into account latent structures in employment decisions across mothers. Finally, my evidence on heterogeneous parental leave effects is especially relevant for policymakers to design efficient parental leave legislation.

The remainder of this paper is structured as follows. Section \ref{Background} explains the institutional background of the parental leave legislation change in Germany and describes the mechanisms through which it can affect the heterogenous employment effects of mothers. Section \ref{Model} introduces the panel structure model and the C-Lasso technique of \cite{Su16}. Section \ref{Data} describes the data set employed in this study. Section \ref{Results} reports the main findings and Section \ref{Conclusion} concludes the paper. The main tables are stated in the Appendix. Readers who are interested in additional insights on the data's sample restriction and the methodology are referred to the Online Appendix.

\section{Institutional background}\label{Background}
\subsection{The German parental benefit reform}
In Germany, mothers have been entitled to government-provided paid leave six weeks before and eight weeks after childbirth since the mid-1950s, and parental leave regulations have continuously extended in a series of reforms starting in the late 1970s.\footnote{For an overview of the developments of the parental leave regulations in Germany, see \cite{Kluv18}, \cite{Kluve2013}, \cite{R19} and \cite{Schoenberg2014}.} Since 1979, mothers have been granted six months of post-birth job protection, and it has been expanded up to a maximum of 36 months after birth since 1992. During this period, government-provided financial transfers have been extended to two years while on leave and targeted to low-income families. In 2007, however, the German parental benefit reform came into force that compensated women much more generous for forgone earnings by taking into account the opportunity costs of child-rearing.

The previous benefit program, `child-raising allowance' (\textit{Erziehungsgeld}), was targeted to low-income households and was much less generous. It provided two options: Option 1 comprised a maximum of 300 EUR per month for up to 24 months after birth (total 7200 EUR), and Option 2 comprises a monthly payment of 450 EUR over 12 months after birth (total 5400 EUR). Benefit eligibility under both options was means-tested on household income, and the eligibility criteria did not permit to work more than 30 h per week during benefit receipt. Of the parents entitled to the parental allowance, the majority (66\%) chose Option 1, and only 10\% chose Option 2.\footnote{24\% of parents were not eligible at all (\cite{Kluve2013}).} 
\begin{table*}[h!]
	\begin{center}
		\small
		\caption{\small Overview over Changes in the Benefit Structure Induced by the Reform}\label{table1}
		\begin{tabular}{p{2.6cm}  >{\raggedleft}p{5cm} p{0.1cm} >{\raggedleft}p{1.6cm} p{1.6cm}<{\raggedleft} }  \hline \\[-0.5em] 
			&Parental money (Post-2007 benefit)&  &\multicolumn{2}{c}{Child-rearing allowance (Pre-2007 benefit)}  \\ \cline{2-2} \cline{4-5} \\[-0.5em] 
			& & & Option 1 & Option 2 \\ \hline 	\\[-0.5em] 	
			Monthly benefit	&  67\% of ave. monthly pre-birth net income; min. 300 EUR, max. 1800 EUR. Mothers with no pre-birth income entitled to 300 EUR.	& &   300 EUR & 450 EUR             \\[0,2cm]
			Means tested & no & & yes & yes                     \\[0,2cm]
			Max. duration & 14 months (2 month for the father)  & & 24 months & 12 months              \\[0,2cm]
			Proportion of parents covered & almost 100\% & & 66\%  & 10\%      \\[0,2cm]
			Requirements &not working more than 30 h per week during receipt  & & \multicolumn{2}{l}{not working more than 30 h per week during receipt}                  \\[0,2cm]
			Total max. benefit & 3,600-21,600 EUR (+ 600-3,600 EUR for the father) & & 7,200 EUR & 5,400 EUR       \\ \hline \\[-1.0em]
			\multicolumn{5}{l}{\small \textit{Notes}: See \cite{R19} for further details of the reform.}	
		\end{tabular}
	\end{center}
\end{table*}
On January 1, 2007, a new parental leave benefit, `parental money' (\textit{Elterngeld}), replaced the previous benefit regulation. Table \ref{table1} summarizes the changes in the benefits system induced by the new parental benefit reform. The objective of the reform is ``to promote the effective and sustainable economic security of families after childbirth through a parental allowance: to avoid income drops, $\dots$ to promote the economic independence of both parents, and to allow a fair compensation of opportunity costs of childbearing'' \cite{BMFSFJ2008}. Whereas the old benefit scheme was means-tested and thus targeted to low-income families, the coverage of the new `parental money' is effectively 100\% of parents. Most importantly, the `parental money' is an earnings-dependent leave benefit, which compensates women according to their childbearing opportunity costs. In contrast to the old means-tested benefits, the new regulation is much more generous: For most mothers, the new benefit replaces 67\% of pre-child birth net labor earnings for up to 14 months after childbirth (12 months if one parent only takes up the benefit). The maximum amount is truncated at 1,800 EUR per month for households at the top of the earnings distribution, and a flat minimum of 300 EUR per month is paid to parents with no previous labor earnings. The monthly benefits translate into a total benefit ranging between 3,600 EUR and 21,000 EUR, depending on where the household is on the earnings distribution. Like the old regulation, benefit eligibility depends on not working more than 30 h a week during benefit receipt. Mothers who choose to work below 30 h during their benefit entitlement face a reduction in benefits with increasing labor earnings.
\subsection{Mechanisms: the heterogeneous effects of the reform on employment}
The current literature on the parental benefit reform's employment effects documents heterogeneous patterns after childbirth based on observables. Whereas previous studies confirm that the reform encourages mothers to remain at home during the first year following childbirth and thus reduces maternal employment during benefit receipt, especially for groups who benefit strongly (e.g., \cite{Ber2011}, \cite{Greyer2015}, \cite{Kluve2013}, \cite{Kluv18}), the study of Kluve and Schmitz finds pronounced beneficial patterns for the medium run (2-5 years after childbirth). In particular, the authors provide evidence that the effects on mothers' employment probability are positive, significant, and large, ranging up to 10\%. 
	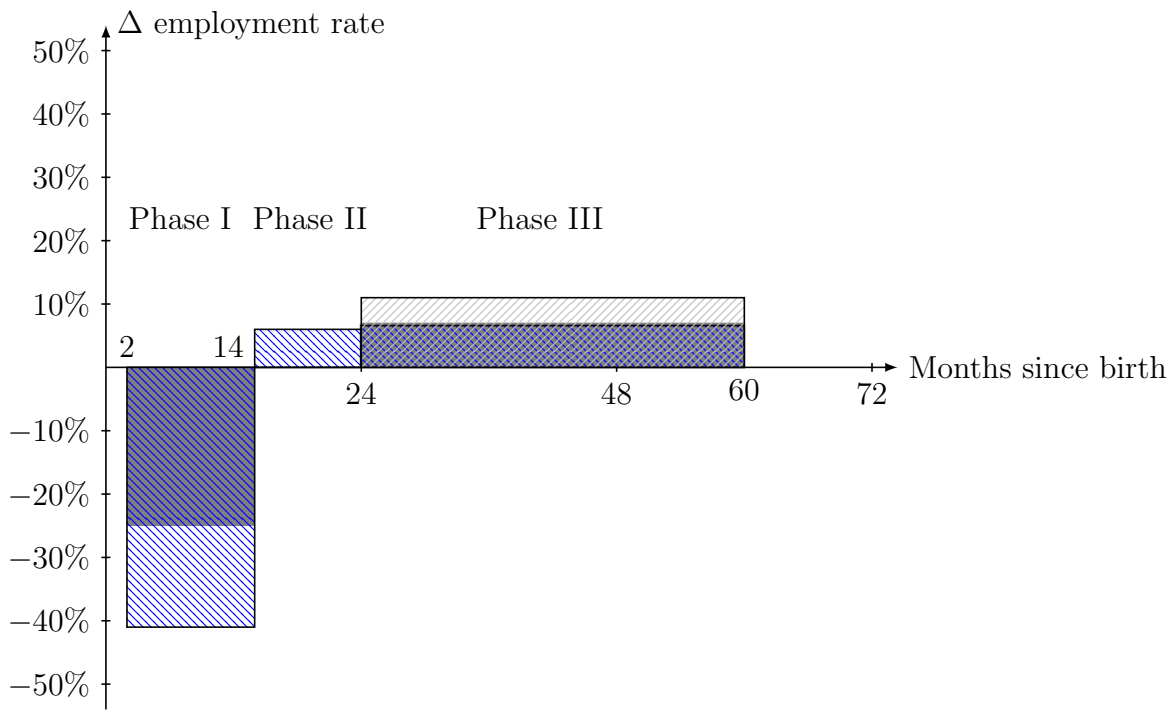
\begin{figure*}
		\captionof{figure}{Recent Findings of the Literature} \label{fig:221}	
		\begin{tikzpicture}[>=latex,semithick, scale = 0.84]	
		\draw[->] (0,-5.4)--(0,5.4) node[right] {$\Delta$ employment rate};	
		\draw[->] (0,0)--(12.4,0) node[right] {Months since birth};
		\foreach \x/\xtext in {   4/24, 8/48, 12/72}
		\draw[shift={(\x,0)}] (0pt,2pt) -- (0pt,-2pt) node[below] {$\xtext$};
		\foreach \y/\ytext in {-1/-10\%,-2/-20\%,-3/-30\%,-4/-40\%, -5/-50\%,1/10\%, 2/20\%,3/30\%, 4/40\%, 5/50\%}
		\draw[shift={(0,\y)}] (2pt,0pt) -- (-2pt,0pt) node[left] {$\ytext$};
		\draw  (0,2)  (1.165,2) node[above] {Phase I};
		\draw  (1.165,2) (3.2,2) node[above] {Phase II};
		\draw  (1.165,2)  (6.8,2) node[above] {Phase III};
		\draw[very thin, color = black ] (2.33,0) node[above left]{14};
		\draw[very thin, color = black] (0.33,0) node[above ]{2};
		\draw[ thick, color = gray] (2.33,0)--(4,0);
		\draw[very thick, color = gray] (4,0)--(4,0.68);
		\draw[very thick, color = gray] (4,0.68)--(10,0.68);
		\draw[very thick, color = black] (10,0) node[below] {60};
		\fill[gray] (0.33,0) -- (0.33, -2.5) -- (2.33, -2.5) -- (2.33, 0);
		\fill[gray] (4,0) -- (4, 0.68) -- (10, 0.68) -- (10, 0);
		\draw[pattern=north west lines, pattern color=blue] (0.33,0) rectangle (2.33,-4.1);
		\draw[pattern=north west lines, pattern color=blue] (2.33,0) rectangle (4.0,0.6);
		\draw[pattern=north west lines, pattern color=blue] (4.0,0) rectangle (10,0.66);
		\draw[pattern=north east lines, pattern color=lightgray] (4.0,0) rectangle (10,1.1);
		\end{tikzpicture}
		\captionof*{figure}{\small \textit{Notes}: Changes in Employment Rates induced by the Reform. \\ \textcolor{gray}{Filled gray}: All mothers. \textcolor{blue}{Shaded blue}: Upper tercentile. \textcolor{lightgray}{Shaded lightgray}: Middle tercentile.}	
	\end{figure*}

Given the earnings-dependent design of the reform, quite heterogeneous labor supply reactions of mothers following the birth of a first child can be expected. The existing literature confirms that the reform impact on maternal employment differs across socioeconomic groups (\cite{Kluve2013}, \cite{Kluv18}). Figure \ref{fig:221} summarizes the findings of \cite{Kluv18}, and shows the different changes in employment rates across income terciles induced by the reform. For all mothers, the authors report a 25\% decrease in maternal employment over the control mean during Phase I, mainly driven by mothers from the upper tercile of the income distribution (41\%) and first-time mothers (32\%, not reported in Fig. \ref{fig:221}), which are the groups most strongly incentivized by the reform. In Phase II, they find no significant impacts on employment for most socioeconomic groups. Only treatment group mothers in the upper tercile of the income distribution display a 6\% increase over the control mean. For Phase III, the authors find that the reform substantially increased maternal employment, especially for mothers from the middle and upper terciles of the income distribution, partially offsetting the negative short-run employment effects. 

In sum, the literature on reform's employment effects documents heterogeneous patterns after birth concerning observables (mostly across income and education groups). However, the 2007 reform in paid leave may affect employment decisions also concerning unobservables: e.g., mothers who want to stay longer at home to spend more time with their child - and pre-reform were not able to take up longer leave because of financial constraints - may decrease their labor supply under the more generous regulation. On the other hand, as paid leave benefits are paid conditional on pre-birth earnings, the reform may positively impact employment. E.g., Career-oriented mothers may directly return to work with the end of the benefit receipt to smooth the household's income and minimize childraising opportunity costs.

\section{Model and Methodology}\label{Model}
In this section, I formulate the panel structure model and describe the methodology introduced by \cite{Su16}.
\subsection{Model}
I consider data in form of a real-valued deterministic design $\boldsymbol{X} = [\boldsymbol{X}_{11} \ \boldsymbol{X}_{12}]\in \mathbb{R}^{(N T) \times p} $ and a binary response $\boldsymbol{y} \in \mathbb{R}^{N T}$. The design $\boldsymbol{X}$ is partitioned in two blockmatrices $\boldsymbol{X}_{11} \in \mathbb{R}^{(N T) \times q}$ and $\boldsymbol{X}_{12} \in \mathbb{R}^{(N T) \times (p-q)}$, with $q \le p$. I denote the rows of $\boldsymbol{X}$ by $\boldsymbol{x}_{it} \in \mathbb{R}^p$ where $i=1,\dots, N, t=1, \dots,T$, and the columns of $\boldsymbol{X}$ by $\boldsymbol{x}^j \in \mathbb{R}^{N T}$ where $j = 1,\dots p$. Similarly, I denote the rows of $\boldsymbol{X}_{11}$ and $\boldsymbol{X}_{12}$ by $\boldsymbol{x}_{11_{it}} \in \mathbb{R}^q$ and $\boldsymbol{x}_{12_{it}} \in \mathbb{R}^{p-q}$, respectively. The columns of $\boldsymbol{X}_{11}$ and $\boldsymbol{X}_{12}$ are denoted by  $\boldsymbol{x}_{11}^{j_1} \in \mathbb{R}^{NT}$ and $\boldsymbol{x}_{12}^{j_2} \in \mathbb{R}^{NT}$ with $j_1=1,\dots,q$ and $j_2=q+1,\dots,p$, respectively.  \\
The econometric model takes the form of a dynamic binary choice model
\begin{align}
y_{it} = \mathbf{1} \{\boldsymbol{x}_{11_{it}}^{\top}\boldsymbol{\gamma^*} + \boldsymbol{x}_{12_{it}}^{\top} \boldsymbol{\beta^*}_i+ \mu_i^*- \epsilon_{it} > 0 \}, \quad i=1,\dots,N; t=1, \dots, T,
\end{align}
where $i$ and $t$ denote mother $i$ and month after childbirth $t$, respectively, $\boldsymbol{x}_{it} =[\boldsymbol{x}_{11_{it}} \ \boldsymbol{x}_{12_{it}} ] \in \mathbb{R}^p $ is the vector of covariates containing also the dynamic component $y_{i,t-1}$, $\boldsymbol{\gamma^*} \in \mathbb{R}^q$ is the unknown regression vector that is common across all $i$, $\boldsymbol{\beta^*}_i \in \mathbb{R}^{p-q}$ is the unknown individual specific regression vector, $\mu_i^*$ denotes the unknown individual fixed effect, and $\epsilon_{it}$ is the idiosyncratic error term. Note, when $q=0$, all parameters of interest are assumed to be individual-specific, and when $q=p$, all parameters of interest are assumed to be common across $i$. In case of $0<q<p$, I refer to the mixed panel structure model of \cite{Su16} where some parameters of interest are common across all individuals whereas others are individual-specific.

A latent group structure is imposed on the $\boldsymbol{\beta^*}_i$. For this purpose, I consider a partition of the index set $\{1,\dots,N\} $, that is, a collection of disjoint sets $\mathcal{G}^1, \dots, \mathcal{G}^{K^*}$ that satisfy $\cup_{k=1}^{K^*} \mathcal{G}^k =\{1,\dots,N\}$. $K^*$ denotes the true a priori unknown number of groups. In particular, I assume for $\boldsymbol{\beta^*}_i$ a \textit{sparsity} inducing group structure of the following form
\begin{align}
\boldsymbol{\beta^*}_i = \left\{
\begin{array}{ll}
\boldsymbol{\alpha^*}_{\mathcal{G}^1} & \quad \quad \text{if} \ i \in \mathcal{G}^1 \\
\vdots & \quad \quad \quad \ \ \vdots \\
\boldsymbol{\alpha^*}_{\mathcal{G}^{K^*}} & \quad \quad \text{if} \ i \in \mathcal{G}^{K^*}, 
\end{array}
\right.
\end{align}
where $K^*\ll N $ and $\boldsymbol{\alpha^*}_{\mathcal{G}^k} \neq \boldsymbol{\alpha^*}_{\mathcal{G}^l}$ for $k \neq l$. Intuitively, the above model says that mothers in the same Group $\mathcal{G}^k$ share the same parameter vector $\boldsymbol{\alpha^*}_{\mathcal{G}^k}$, and mothers in different groups have parameter vectors that differ from each other.
\subsection{Methodology}
Following \cite{Su16}, the goal is to obtain \textit{classifier-Lasso} (C-Lasso) estimates $\boldsymbol{\hat{\gamma}}$, $\boldsymbol{\hat{\beta}}_i$, $\boldsymbol{\hat{\alpha}}_{\hat{\mathcal{G}}^k}$ with the family of regularized estimators
\begin{align}\label{eq:21}
\boldsymbol{\hat{\gamma}}, \boldsymbol{\hat{\beta}}_i, \boldsymbol{\hat{\alpha}}_{\hat{\mathcal{G}}^k} \in \argmin_{\substack{\boldsymbol{\gamma} \in \mathbb{R}^q, \\ \boldsymbol{\beta}_i, \boldsymbol{\alpha}_{\mathcal{G}^k} \in \mathbb{R}^{p-q}}}\{ Q(\boldsymbol{\gamma},\boldsymbol{\beta}_i) + rh[\boldsymbol{\beta}_i, \boldsymbol{\alpha}_{\mathcal{G}^k}]  \}, 
\end{align}
for an objective function $Q(\cdot)$, a tuning parameter $r \in [0,\infty)$, and with prior function $h: \mathbb{R}^{p-q} \rightarrow [0,\infty]$ which is specified here as
\begin{align}\label{eq:22}
h[\boldsymbol{\beta}_i, \boldsymbol{\alpha}_{\mathcal{G}^k}] := \frac{1}{N} \sum_{i=1}^N \prod_{k=1}^{K} ||\boldsymbol{\beta}_i - \boldsymbol{\alpha}_{\mathcal{G}^k} ||_2,
\end{align}
with the remainder of the notation being defined as above. Following \cite{Su16}, I use fixed effects penalized profile likelihood (PPL) to estimate the unknown parameter vectors in eq. \ref{eq:21}. Therefore, the first term of eq. \ref{eq:21}
\begin{align}\label{eq:23}
Q(\boldsymbol{\gamma}, \boldsymbol{\beta}_i) = &\frac{1}{NT} \sum_{i=1}^N \sum_{t=1}^T y_{it} \log \Phi(y_{it}-\boldsymbol{x}_{11_{it}}^{\top}\boldsymbol{\gamma} -\boldsymbol{x}_{12_{it}}^{\top} \boldsymbol{\beta}_i-\hat{\mu}_i) \nonumber \\ & + (1-y_{it}) \log(1-\Phi(y_{it}-\boldsymbol{x}_{11_{it}}^{\top}\boldsymbol{\gamma} -\boldsymbol{x}_{12_{it}}^{\top} \boldsymbol{\beta}_i-\hat{\mu}_i))
\end{align}
is the negative profile log-likelihood function and $\Phi(\cdot)$ denotes the conditional normal cumulative distribution function (CDF).\footnote{ The individual fixed effect estimates are obtained by
	\begin{align*}
	\hat{\mu_i} \in  \underset{\mu_i \in \mathbb{R}}{\arg \min}  &\frac{1}{T} \sum_{t=1}^T y_{it} \log \Phi(y_{it}-\boldsymbol{x}_{11_{it}}^{\top}\boldsymbol{\gamma}-\boldsymbol{x}_{12_{it}}^{\top} \boldsymbol{\beta}_i-\mu_i) + (1-y_{it})\log(1-\Phi(y_{it}-\boldsymbol{x}_{11_{it}}^{\top}\boldsymbol{\gamma}-\boldsymbol{x}_{12_{it}}^{\top}\boldsymbol{\beta}_i-\mu_i)).
	\end{align*}}
Based on the estimated group structure $\hat{\mathcal{G}}^1, \dots, \hat{\mathcal{G}}^K$ in eq. \ref{eq:21} and \ref{eq:22}, post C-Lasso estimates $\boldsymbol{\hat{\hat{\gamma}}}=(\boldsymbol{\hat{\hat{\gamma}}}_1^{\top}, \dots,\boldsymbol{\hat{\hat{\gamma}}}_K^{\top})^{\top} \in \mathbb{R}^{qK}$,\footnote{Note that each $\boldsymbol{\hat{\hat{\gamma}}}_k$ contains the same estimates because $\boldsymbol{\hat{\hat{\gamma}}}_k$ is not group-specific.} $\boldsymbol{\hat{\hat{\beta}}}=(\boldsymbol{\hat{\hat{\beta}}}_1^{\top}, \dots,\boldsymbol{\hat{\hat{\beta}}}_N^{\top})^{\top} \in \mathbb{R}^{(p-q)N}$ and  $\boldsymbol{\hat{\hat{\alpha}}}= (\boldsymbol{\hat{\hat{\alpha}}}_{\hat{\mathcal{G}}^1}^{\top}, \dots,\boldsymbol{\hat{\hat{\alpha}}}_{\hat{\mathcal{G}}^K}^{\top})^{\top} \in \mathbb{R}^{(p-q)K}$ are obtained. In particular, I use fixed effects quasi-maximum likelihood to estimate the common slope parameters $\boldsymbol{\hat{\hat{\alpha}}}_{\hat{\mathcal{G}}^k}$ and set $\boldsymbol{\hat{\hat{\beta}}}_i=\boldsymbol{\hat{\hat{\alpha}}}_{\hat{\mathcal{G}}^k}$ for all $i \in \hat{\mathcal{G}}^k$ with $\hat{\mathcal{G}}^k:=\{i \in \{1,\dots, N\}: \boldsymbol{\hat{\beta}}_i = \boldsymbol{\hat{\alpha}}_{\hat{\mathcal{G}^k}}\} $ for $k=1,\dots,K$.

As the true number of groups in eq. \ref{eq:22} is a priori unknown, I follow \cite{Su16} and make the dependence of $\boldsymbol{\hat{\beta}}_i$ and $\boldsymbol{\hat{\alpha}}_{\hat{\mathcal{G}}^k}$ on $[K,r]$ explicit. I choose $K$ and $r$ in eq. \ref{eq:21} and \ref{eq:22} to minimize the following BIC-type information criterion
\begin{align}\label{eq:24}
\hat{K}[r] \in \underset{1 \le K \le K_{\max}}{\arg \min}\{ \tilde{Q}(\boldsymbol{\hat{\hat{\alpha}}}_{\hat{\mathcal{G}}^k[K,r]}) + spK  \},            
\end{align}
where $r$, $p$ and $K$ are defined as above, $\boldsymbol{\hat{\hat{\alpha}}}_{\hat{\mathcal{G}}^k[K,r]}= [\boldsymbol{\hat{\hat{\gamma}}}_k \ \boldsymbol{\hat{\hat{\alpha}}}_{\hat{\mathcal{G}}^k}] \in \mathbb{R}^p$ denotes the vector of post C-Lasso estimates, and $s$ is the tuning parameter with $s:=1/4\log(\log(T))/T$. I assume that the true number of groups $K^*$ is bounded from above by $K_{\max}=4$. The first term of eq. \ref{eq:24}
\begin{align}
\tilde{Q}(\boldsymbol{\hat{\hat{\alpha}}}_{\hat{\mathcal{G}}^k[K,r]}) &= \frac{2}{NT}\sum_{k=1}^K \sum_{i \in \hat{\mathcal{G}}^k[K,r]} \sum_{t=1}^T y_{it} \log \Phi(y_{it}-\boldsymbol{x}_{it}^{\top} \boldsymbol{\hat{\hat{\alpha}}}_{\hat{\mathcal{G}}^k[K,r]}-\hat{\mu}_i^{\hat{\mathcal{G}}^k[K,r]}) \nonumber \\ &+ (1-y_{it}) \log(1-\Phi(y_{it}-\boldsymbol{x}_{it}^{\top}\boldsymbol{\hat{\hat{\alpha}}}_{\hat{\mathcal{G}}^k[K,r]}-\hat{\mu}_i^{\hat{\mathcal{G}}^k[K,r]}))
\end{align}
denotes again the neagative profile log-likelihood function and $\hat{\mu}_i^{\hat{\mathcal{G}}^k[K,r]}$ is a group-specific individual fixed effect.

\section{Data}\label{Data}
My analysis draws on a novel custom shaped administrative data set based on social security records made to investigate female employment biographies in Germany.\footnote{This data set (FEMPSO BIRTH) was created by the project 7 (Female Employment Patterns, Fertility, Labor Market Reforms, and Social Norms: A Dynamic Treatment Approach) and project 16 (Custom Shaped Administrative Data for the Analysis of Labour Market (CADAL)) of the priority program 1764 of the German Science Foundation (DFG). FEMPSO BIRTH became available in 2018.} 
\footnote{The same data set (FEMPSO BIRTH) is used in another joint project in \cite{Klose}, where we predict the incidence of future first births based on employment records within a machine learning approach. Reference is also made to this paper for the data.}
It is based on two different data sets: employment records from the Integrated Employment Biographies (IEB) prepared by the Federal Employment Agency's Research Data Centre (RDC) at the Institute for Employment Research and on a subsample of microdata of actively insured persons compiled by the German Federal Pension Insurance's RDC. 
Both data sets stem from the same source, the registration procedure for social insurance. 

The version of the IEB used in this paper comprises all individuals in Germany that contributed to the social security system between 1975 and 2014 or obtained transfer payments from the labor agency or were registered as job seekers.\footnote{Not included are civil servants including the majority of teachers as well as the self-employed.}  
Anonymous weekly subsamples of the data are accessible for scientific research (e.g., the ``SIAB'') and have been used in several studies on the employment of mothers and the effects of maternity leaves (see, e.g., \cite{Adda2017}; \cite{Ejrnaes2013}; \cite{Schoenberg2014}). 
The IEB provide detailed daily information on wages, employment, occupations, periods of unemployment, receipt of unemployment benefit as well as basic sociodemographics. An important advantage of the IEB data is that, because of the administrative data's longitudinal nature, employment histories are measured precisely. The main disadvantage of the IEB data is that it does not contain direct information on childbirths. The data only provides information on whether and when a woman takes leave of absence. This causes potential misspecification errors because not all leave-taking may be due to maternity leave, but due to other reasons like illness. Moreover, the child's birthday has to be approximated as six weeks after the mother takes leave since mandatory maternity leave entitlements cover six weeks before and two months after the due date. This may lead to measurement errors in the child's month of birth. Most importantly, this identification strategy cannot capture births by mothers who are out of the workforce at the timepoint of childbirth.

To address these drawbacks, for the custom-shaped data set, precise information on childbirths from the German Pension Insurance has been merged to the IEB. The pension registry gathers all employed persons in the private and public sectors because for those it is mandatory to make a contribution to the statutory pension insurance.\footnote{Excluded from the data are also the civil servants (including teachers), most self-employed, and the economically inactive because they are not covered by statutory pension insurance.} Most importantly, pension accounts provide a fertility record for any woman who has been ever enrolled in the pension insurance because women receive an automatic contribution for child-rearing years. This is basically equivalent to having appeared in the IEB. The fertility information and some information on employment, wages, and basic sociodemographics are available for a particular sample conducted by the German Pension Insurance's RDC.\footnote{Various data products of the German Pension Insurance's RDC rely on this sample, which is referred to as the ``Versicherungskontenstichprobe''. It is a representative stratified sample from all actively insured individuals, drawn in 1983 and kept up and complemented with younger cohorts since then. See, e.g., \citealp{Kreyenfeld2008}.} The merge of the two data sets relies on statistical matching principles since merging based on social security numbers or names was not possible due to the unavailability in the RDC at the Institute for Employment Research. For each woman in the pension data, a statistical twin has been identified in the IEB data based on her year and month of birth, regional information, and employment records. To achieve an accurate match, the data is limited to women without a GDR-employment-biography and with a certain labor market attachment.\footnote{Readers who are interested in additional details on the match of the two data sets are referred to the Online Appendix of \cite{Klose}.} In this paper, I only focus on the matches that are almost certainly assessed to be correct. The resulting data set covers information on around 56\,000 mothers and the years 1975 and 2014.

The sample is further restricted to mothers aged at least 17 before giving birth to their first child. Before age 17, many mothers have not yet entered the labor market because they are still in education and thus have no employment records. Finally, to identify the reform effect, I consider mothers with a first child born between October 1 and December 31, 2006 (control sample), and mothers with children born between January 1 and March 31, 2007 (treatment sample), respectively. The overall number of observations available for the analysis equals 785 mothers. The treatment group comprises 387 mothers, and the control group has 398 mothers. The final data set is a balanced monthly panel that collects maternal employment information up to 72 (+ 1 for the month of birth) months after birth. It consists of West German mothers who have some attachment to the labor market and have children born between the last quarter of 2006 and the first quarter of 2007. 

 \begin{table*}[h!]
	\begin{center}
		\small
		\caption{\small Descriptives}\label{table2}
		\begin{tabular}{p{3.5cm}  >{\raggedleft}p{1.2cm} >{\raggedleft}p{1.2cm}  >{\raggedleft}p{1.2cm} >{\raggedleft}p{1.2cm}  p{0.2cm}  >{\raggedleft}p{1.2cm} >{\raggedleft}>{\raggedleft}p{1.2cm}  >{\raggedleft}p{1.2cm} p{1.2cm}<{\raggedleft}}  \hline \\[-0.5em] 
			& \multicolumn{4}{c}{Pre-Reform Sample} &  &\multicolumn{4}{c}{Post-Reform Sample}  \\ \cline{2-5} \cline{7-10} \\[-0.5em] 
			Variable & Mean & Std. Dev. & Min& Max & & Mean & Std. Dev. & Min & Max \\[0,2cm] \hline \\[-0.5em] 		
			Age & 27.98 & 7.07 & 17 & 56 & & 27.83 & 6.98 & 17 & 51      \\[0,2cm]
			Mother's age at $1st$ birth & 30.10 & 4.35 & 19 & 48 & & 30.80 & 4.29 & 20 & 44   \\[0,2cm]    
			Mother's age at $2nd$ birth$^*$ & 32.87 & 3.95 & 22 & 42 & & 32.84 & 4.06 & 22 & 42  \\[0,2cm] 
			Mother's age at $3rd$ birth$^+$ & 34.52 & 3.98 & 23 & 41 & & 34.48 & 3.91 & 26 & 42	\\[0,2cm] \hline \\[-0.5em]
			Employed & 0.4953 & 0.5000 & 0 & 1 & & 0.4928 & 0.5000 & 0 & 1     \\[0,2cm]  
			Full-time employed & 0.3155 & 0.4647 & 0 & 1 & & 0.3152 & 0.4646 & 0 & 1         \\[0,2cm]
			Part-time employed & 0.1194 & 0.3242 & 0 & 1 & & 0.1270 & 0.3330 & 0 & 1      \\[0,2cm]
			Marginally employed & 0.0626 & 0.2423 & 0 & 1 & & 0.0515 & 0.2211 & 0 & 1        \\[0,2cm] 
			Daily wage & 29.07 & 38.96 & 0 & 201.5 & & 29.86 & 39.49 & 0 & 201.5	\\[0,2cm] \hline \\[-0.5em]
			Low educated & 0.1964 & 0.3972 & 0 & 1 & & 0.2203 & 0.4144 & 0 & 1      \\[0,2cm]
			Middle educated & 0.6485 & 0.4775 & 0 & 1 & & 0.6270 & 0.4836 & 0 & 1    \\[0,2cm]
			High educated & 0.1347 & 0.3414 & 0 & 1 & & 0.1305 & 0.3369 & 0 & 1  \\[0,2cm] \hline \\[-1.0em]
			\multicolumn{10}{l}{\small \textit{Notes}: Control sample (2006m10-m12): $N_C=398$. Treatment sample (2007m1-m3): $N_T=387$.}	\\
			\multicolumn{10}{l}{\small Prebirth period included.}	\\
			\multicolumn{10}{l}{\small $^*$ 237 (60\%) and 245 (63\%) mothers have a $2nd$ child in the control and treatment samples, respectively.}\\
			\multicolumn{10}{l}{\small $^+$ 48 (12\%) and 44 (11\%) mothers have a $3rd$ child in the control and treatment samples, respectively.}	
		\end{tabular}
	\end{center}
\end{table*}

Table \ref{table4} provides descriptive statistics for some key variables by sample. The monthly information relates to the pre- and post-birth periods, i.e., the entire available employment biographies of the considered mothers are taken into account. The summary statistics show that no systematic differences -- at least for the covariates that are agnostic to post-reform reactions -- emerge in the sociodemographic characteristics between the pre- and post-reform samples.\footnote{Balancing tests on the pre-birth covariates, similar to those of \cite{Kluv18}, are omitted because the sample size in the present study is much smaller than the mentioned study and thus more prone to random variations, which would vanish with increasing sample size.} This pattern confirms the quasi-random assignment identification strategy suggested by \cite{Kluv18}. In the majority of cases, mothers have a vocational degree (women with a birth in 2006 Q4/2007 Q1: 65\%/63\%), in around 13\%/13\% of cases women with a birth in 2006 Q4/2007 Q1 are tertiary educated, and, in 20\%/22\% of cases, women have no postsecondary degree. The average employment rate over the considered time span is 49\%/50\% for women with a birth in 2006 Q4/2007 Q1. The employment rate is highest for full-time work (women with a birth in 2006 Q4/2007 Q1: 32\%/32\%), followed by part-time work (women with a birth in 2006 Q4/2007 Q1: 12\%/13\%) and marginal employment (women with a birth in 2006 Q4/2007 Q1: 6\%/5\%). The average age in both samples is between 27.8 and 28 years, and between 30.1 and 30.8 years at first birth. 60\%/63\% of the mothers have a second child, and 12\%/11\% of the mothers have a third child in the pre-/post-reform sample. The average daily wage among working and non-working women is around 29 EUR/30 EUR for the pre-/post-reform sample.

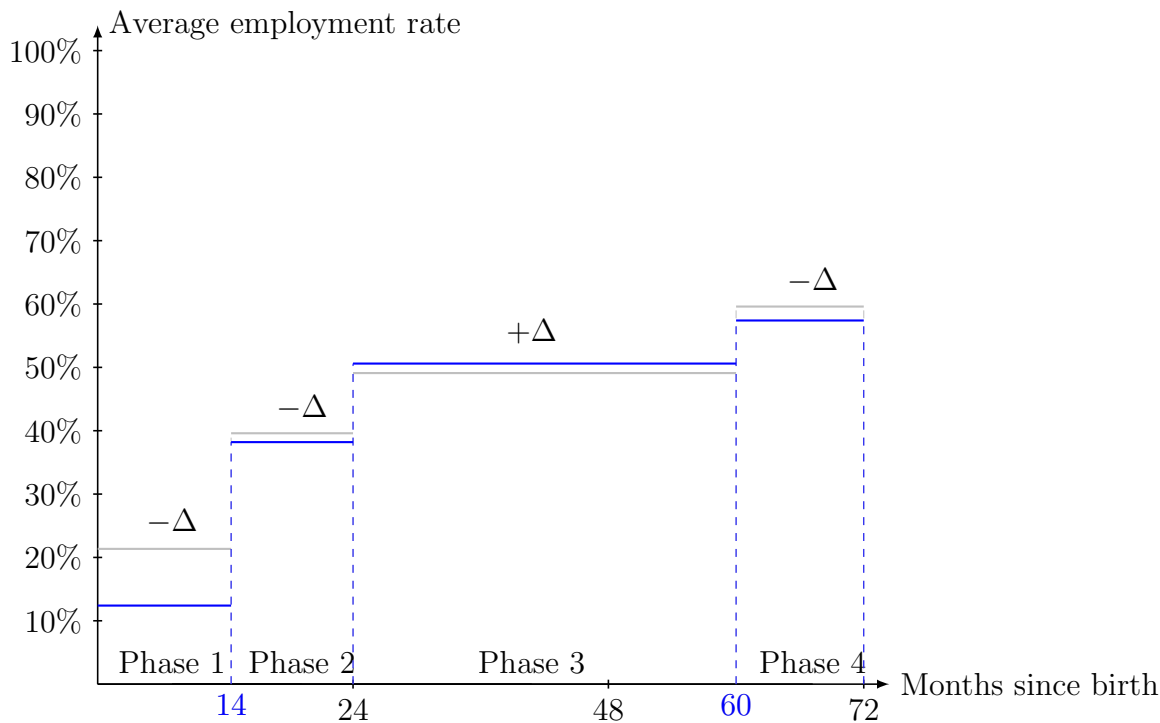
\begin{figure*}
	\captionof{figure}{Changes in the Employment Structure induced by the Reform} \label{fig:1}	
	\begin{tikzpicture}[>=latex,semithick, scale = 0.84]	
	\draw[->] (0,0)--(0,10.4) node[right] {Average employment rate};	
	\draw[->] (0,0)--(12.4,0) node[right] {Months since birth};
	\foreach \x/\xtext in {   4/24, 8/48, 12/72}
	\draw[shift={(\x,0)}] (0pt,2pt) -- (0pt,-2pt) node[below] {$\xtext$};
	\foreach \y/\ytext in {1/10\%, 2/20\%,3/30\%, 4/40\%, 5/50\%, 6/60\%, 7/70\%, 8/80\%, 9/90\%, 10/100\%}
	\draw[shift={(0,\y)}] (2pt,0pt) -- (-2pt,0pt) node[left] {$\ytext$};
	\draw[thick, color = lightgray] (0,2.135) --(2.09,2.135) ;
	\draw  (0,0) -- (1.165,0) node[above] {Phase 1};
	\draw  (1.165,0) -- (3.2,0) node[above] {Phase 2};
	\draw  (1.165,0) -- (6.8,0) node[above] {Phase 3};
	\draw  (1.165,0) -- (11.2,0) node[above] {Phase 4};
	\draw[very thin, color = lightgray, dashed] (2.09,0) node[below] {14} --(2.09,3.960);
	\draw[ thick, color = lightgray] (2.09,3.960)--(4,3.960);
	\draw[very thin, color = lightgray, dashed] (4,0)--(4,4.910);
	\draw[ thick, color = lightgray] (4,4.910)--(10,4.910);
	\draw[very thin, color = lightgray, dashed] (10,0) node[below] {60} --(10,5.960);
	\draw[thick, color = lightgray] (10, 5.960)--(12, 5.960);
	\draw[very thin, color = lightgray, dashed] (12,0)--(12,5.960);
	
	\draw[thick, color = blue] (0,1.240) --(2.09,1.240) ;
	\draw[ thick, color = blue] (2.09,3.820)--(4,3.820);
	\draw[ thick, color = blue] (4,5.060)--(10,5.060);
	\draw[thick, color = blue] (10, 5.74)--(12, 5.74);
	\draw[very thin, color = blue, dashed] (2.09,0) node[below] {14} --(2.09,3.820);
	\draw[very thin, color = blue, dashed] (4,0)--(4,5.060);
	\draw[very thin, color = blue, dashed] (10,0) node[below] {60} --(10,5.74);
	\draw[very thin, color = blue, dashed] (12,0)--(12,5.74);
	
	\draw[thick, color = black]   (1.165,2.20) node[above] {$-\Delta$};
	\draw[thick, color = black]   (3.2,4.00) node[above] {$-\Delta$};
	\draw[thick, color = black]  (6.8,5.20) node[above] {$+\Delta$};
	\draw[thick, color = black]   (11.2, 6.00) node[above] {$-\Delta$};
	\end{tikzpicture}
	\captionof*{figure}{\small \textit{Notes}: \textcolor{lightgray}{Light gray}: Control sample. \textcolor{blue}{Blue}: Treatment sample.}	
\end{figure*}

Figure \ref{fig:1} descriptively summarizes the changes in the employment structure induced by the reform. The $x$-axis indicates the time since childbirth provided in months. Similar to \cite{Kluv18}, four phases emerge: the first phase (from 1 to 14 months) covers the periods of the mandatory maternity leave and the parental benefit receipt. The second phase (up to 24 months) covers the second year after childbirth. The third phase (from 25 to 60 months) covers years three to five after childbirth. The last phase (up to 72 months) comprises the sixth year after childbirth, which denotes the reference group in the estimations later on. The $y$-axis denotes the average employment rate in percent. The employment situations before and after the reform are depicted in light gray and blue, respectively. Figure \ref{fig:1} confirms the evidence from previous research (e.g., \cite{Greyer2015}, \cite{Kluve2013}, \cite{Kluv18}): Phase 1 leads to a significant drop in maternal employment rates because of the presence of the parental benefit, which induces on average a large delta in benefit levels during the first 14 months after childbirth. Phase 2 shows only a slight, almost negligible decrease in the average employment rate of mothers. In Phase 3, however, the average maternal employment rate increases. The last phase again shows a decrease in the average employment rate.

Finally, as a data preprocessing step, I drop all samples from the data that do not have any variation in the dependent variable (either $\frac{1}{72}\sum_{t=0}^{72} y_{it} = 0$ or $\frac{1}{72}\sum_{t=0}^{72} y_{it} = 1$ for $i \in \{1,\dots, 398 \ (387)\}$), as they would lead to infinite ($\pm \infty$) fixed effects in the fixed effects probit model considered here.\footnote{Refer to the Online Appendix for more information on this specification problem.} Overall, this results in sample sizes of $N_C = 329$ (control group) and $N_T = 333$ (treatment group) available for the analysis.

\section{Empirical Results}\label{Results}
As previously explained, I provide estimates of the heterogeneous employment effects up to six years after childbirth by exploiting the latent group structures in mothers' return behavior to employment. Moreover, I estimate the policy impacts of the parental benefit within the different groups of mothers by identifying the reform effect similar to \cite{Kluv18}. Before turning to my main analysis, however, I first introduce the empirical approach, and in a second step, I also report empirical estimates from the traditional homogenous panel data model similar to that reported by \cite{Hys99}.    

\subsection{Empirical approach}
To investigate maternal employment after the first birth, I split the child age effect into four phases. The first phase comprises the period of mandatory maternal protection immediately after childbirth (up to 2 months after birth) and the parental benefit receipt (3.-14. month after birth). The second phase (up to 24 months) covers the second year after childbirth and is the short-run perspective. The medium-run perspective contains years three to five after childbirth and denotes the third phase. The fourth phase covers the sixth year after childbirth. 

The empirical benchmark model has the following form:
\begin{align}\label{eq:1}
\text{employ}_{it} &=  \mathbf{1}\{\alpha \ \text{employ}_{i,t-1} + \beta \ \text{child}_{\text{2nd}_{it}} + \gamma \ \text{child}_{\text{m1-m14}_{it}} +   \delta \ \text{child}_{\text{m15-m24}_{it}}  + \eta \ \text{child}_{\text{m25-m59}_{it}} \nonumber \\ &+ \zeta_i - \epsilon_{it}  \ge 0     \},
\end{align} 
where $employ_{it}$ is the employment indicator, $employ_{i,t-1}$ is the employment indicator of the previous period, $child_{2nd_{it}}$ is a dummy variable controlling for a second child, $child_{m1-m14_{it}}$, $child_{m15-m24_{it}}$ and $child_{m25-m59_{it}}$ are the child age indicators for the first child, $\zeta_i$ is an individual fixed effect and $\epsilon_{it}$ is an idiosyncratic error term. $child_{m60-m72_{it}}$ is the omitted reference group. The three child age indicators' estimates measure the effect during the benefit receipt, the short-run effect immediately after benefit exhaustion, and the medium-run effect three to five years after childbirth, respectively. I do not include time-constant covariates such as education level as they are absorbed in the fixed effects.
By combining this baseline specification with the panel structure formulation of \cite{Su16} that allows individual-specific coefficients, I consider the following two specifications: 
\begin{align}\label{eq:2}
\text{employ}_{it} &=  \mathbf{1}\{\alpha \ \text{employ}_{i,t-1} + \beta \ \text{child}_{\text{2nd}_{it}} + \gamma_i \ \text{child}_{\text{m1-m14}_{it}} +  \delta_i \ \text{child}_{\text{m15-m24}_{it}}  + \eta_i \ \text{child}_{\text{m25-m59}_{it}} \nonumber \\  &+ \zeta_i - \epsilon_{it}  \ge 0     \}, 
\end{align} 
and
\begin{align}\label{eq:3}
\text{employ}_{it} &=  \mathbf{1}\{\alpha_i \ \text{employ}_{i,t-1} + \beta_i \ \text{child}_{\text{2nd}_{it}} + \gamma_i \ \text{child}_{\text{m1-m14}_{it}} +  \delta_i \ \text{child}_{\text{m15-m24}_{it}}  + \eta_i \ \text{child}_{\text{m25-m59}_{it}}  \nonumber \\  &+ \zeta_i - \epsilon_{it}  \ge 0     \}, 
\end{align} 
which allow for individual specific slope coefficients ($\alpha_i,\beta_i, \gamma_i, \delta_i, \eta_i$). Following \cite{Su16}, I allow the parameters ($\gamma_i, \delta_i, \eta_i$)  in \ref{eq:2} and ($\alpha_i,\beta_i, \gamma_i, \delta_i, \eta_i$) in \ref{eq:3} to follow certain latent group structures. Equation \ref{eq:2} is an extension to mixed panel structure models, where the parameters of lagged employment and the second child are constrained to be common across all mothers, whereas the child age parameters of the first child are assumed to be group-specific.

To identify the policy impacts of the parental leave reform for all mothers and the identified groups, I use a design idea that builds on previous work (\cite{Kluv18}, \cite{Kluve2013}). \cite{Kluv18} adopt a sharp RD design that compares the subsequent labor supply behavior of mothers who gave childbirth shortly before the reform with mothers who gave birth shortly after the reform. Because the authors argue that the quasi-random assignment to treatment and control groups is achieved within a range of three months before and after the discontinuity point of the 1\ts{st} January 2007, I consider all mothers with first children born between the 1\ts{st} October 2006 and 31\ts{st} March 2007. Mothers with children born between the 1\ts{st} October and 31st December 2006 belong to the control sample, and mothers with children born between the 1\ts{st} January and 31st March 2007 belong to the treatment sample. 

Perhaps the most straightforward way to model the reform effect within a sharp regression discontinuity design is to include an indicator for the treatment group and to run a linear regression on the entire sample. The treatment indicator captures the average causal impact of the reform on employment. However, in the context of a fixed effect approach, it is not possible to include time-constant covariates as they are absorbed in the fixed effects. Since the entitlement to parental allowance is constant over time (mothers are either eligible or not), such a modeling strategy seems inappropriate.\footnote{One possibility to solve this problem is the interaction of the reform indicator with the child age indicators. The interaction terms capture the employment effects of the reform within the different child age phases.} Therefore, I estimate the reform effect indirectly by separately running C-Lasso estimation for the treatment and control samples. Although I cannot provide direct causal evidence on how an increase in paid parental leave for a post-reform baby can affect maternal employment, I can apply an identification strategy that compares the treatment and control samples in terms of differences in the average predicted employment probabilities. To exploit these differences, I compute predictions of the average employment probabilities: 
\begin{align}
\overline{\mbox{$\text{employ}$\raisebox{2mm}{}}}_j = 1/(N[\hat{\mathcal{G}}^k] T[p^j]) \ G(\boldsymbol{x}_{i[\hat{\mathcal{G}}^k]t[p^j]}^{\top} \boldsymbol{\hat{\hat{\alpha}}}_{\hat{\mathcal{G}}^k}+\hat{\mu}_i[\hat{\mathcal{G}}^k]),
\end{align}
where $G(\cdot)$ denotes the CDF of the standard normal, $p^j, \ j=1,\dots, 4$ denotes the four different phases of child age, $[\cdot]$ denotes the restriction to the group-specific subspace, and the remainder is defined as in the methodology section.\footnote{Note that for computing the predictions of the average employment probabilities, I do not take into account the effect of the second child.} Then, I take the differences in the control- and treatment-specific average employment probability predictions. These differences $\Delta_{\text{T-C}_j}$ capture the impact of the reform on maternal employment within the different child age phases. Finally, I test if the differences in the predicted probabilities are statistically different from zero:\footnote{cf. \cite{Long} for performing group comparisons in nonlinear models using predicted probabilities.}
\begin{eqnarray}\label{eq:4}
H_0: \frac{\Delta_{T-C}}{\sqrt{\frac{\sigma^2_T}{N[\hat{\mathcal{G}}^k]_T} +\frac{\sigma^2_C}{N[\hat{\mathcal{G}}^k]_C}}}=0,
\end{eqnarray}
where the plug-in principle is used to estimate the variances of the predicted probabilities. Since I test the employment probability predictions of two independent samples, the covariance term drops out in the denominator of eq. \ref{eq:4}.

The advantage of this identification strategy is that I can disentangle the latent effect heterogeneity of the reform utilizing data-determined grouping. To test the differences in the predicted employment probabilities of the respective control and treatment pairs and to give them an interpretation in causal direction, strong assumptions are required. First, it is assumed that the C-Lasso identifies the same number of groups for the control and treatment samples. Second, it is assumed that the C-Lasso identifies the same drivers of unobserved heterogeneity in the respective control and treatment pairs. Only in this case, the respective control group constitutes a valid counterfactual situation for the treatment group. Finally, it is assumed that the statistically significant differences in the predicted employment probabilities can be interpreted in a causal direction. In the following sections, I work under these assumptions. Moreover, in the Online Appendix, I develop a distance-based criterion that should identify the associated control and treatment pairs under which the assumptions are most likely to hold. 
\subsection{Baseline estimation}
Before reporting my estimation results of Eq. \ref{eq:1}, in Table \ref{table3}, I provide baseline estimates for maternal employment rates from the standard Linear Probability Model (LPM) with fixed effects.\footnote{In the Appendix in Table \ref{table33}, I also provide baseline estimates from standard LPM without fixed effects.} 
\vspace{0,4cm}
\begin{center}
	[Insert Table \ref{table3} here.]
\end{center}
\vspace{0,4cm}
As columns (1), (2), (4), and (5) of Table \ref{table3} show, the results for the Phases I, II, and III show the same pattern identified in previous research (e.g., \cite{Kluv18}, \citealp{Schoenberg2014}): once for individual-specific effects, lagged employment and a potential second child is controlled, mothers in the treatment sample display a statistically significant lower average probability of being employed during the first 14 months after childbirth compared to the control sample.\footnote{Testing H$_0$: child$_{{\text{m1-m14}_{it}}_T}-$ child$_{{\text{m1-m14}_{it}}_C}=0$ results in a small $p$-value. cf. columns (4) and (5) of the lower part of Table \ref{table3}.} Indeed, having a child aged between 1 and 14 months instead of a child aged between 60 and 72 months reduces the probability of maternal employment on average by 11.4 percentage points for the treatment sample holding constant all other factors. In contrast, the control sample displays an 8.4 percentage point lower employment probability on average. For Phases II and III, I cannot find statistically significant differences between the control and treatment samples. In fact, for mothers of the treatment sample, having a child aged between 25 and 59 months instead of a child aged between 60 and 72 months reduces the probability of employment on average by 2.2 percentage points holding constant all other factors. I find a 2.3 percentage points lower employment probability on average for corresponding mothers of the control sample. For an `in-group' comparison, I additionally test whether the three child age estimates within the pre-/post-reform sample are statistically different from each other. In the upper part of Table \ref{table3}, columns (3)-(5) provide the null hypothesis and the respective $p$-value for testing the linear combinations. The $p$-values show that all estimates are statistically different from each other. Although my results confirm the evidence from previous studies in the short run, I do not get evidence of pronounced beneficial medium-run labor market effects of the parental benefit reform identified by \cite{Kluv18}. 

Next, I estimate Eq. \ref{eq:1} and report the results for the coefficient estimates and the corresponding marginal effects in columns (1)-(2) and (3)-(4) of Tables \ref{table9} and \ref{table4}.
\vspace{0,4cm}
\begin{center}
	[Insert Table \ref{table9} and Table \ref{table4} here.]
\end{center}
\vspace{0,4cm}
Columns (1) and (2) of Table \ref{table9} provide coefficient estimates from a dynamic probit FE model without bias correction. The estimates in columns (3) and (4) are bias-corrected by the half-panel jackknife \cite{Dhaen15}. All standard errors are clustered at the individual level. In line with the evidence from previous research (e.g., \cite{Car01}, \cite{Hys99},\cite{Mich11}), employment is highly persistent, and its association with the child age effects remains robust across LPM FE, Probit FE, and Probit FE JK. For both the control and treatment samples, the marginal effects on lagged employment show that mothers, who were employed the previous month instead of not employed, are, on average, about 80 percentage points more likely to be employed this month when holding constant all other factors. The marginal effects derived from the FE probit estimates are all statistically different from zero and similar to those obtained by classical LPM with fixed effects. My results for the control sample show that having a first child in Phases I, II, and III instead of having a first child in Phase IV decreases maternal employment probability on average by 13.8, 8.8, and 3.7 percentage points holding constant all other factors. The corresponding estimates for the treatment sample are -18.1, -9.0, and -3.7. The $t$-statistics in column (1) of Table \ref{table99} also show that all post-Lasso-PPL estimates of the child age phases are statistically different from each other.\vspace{0,4cm}
\begin{center}
	[Insert Table \ref{table99} here.]
\end{center}
\vspace{0,4cm}
Besides, I test whether the differences in the child age effects for the treatment and control samples are statistically different from zero. Conducting tests for group comparisons like in linear regressions can lead to incorrect conclusions in nonlinear models such as probit since they mix the regression coefficients with the residual variation \cite{All1999}. \cite{All1999} suggests a test that removes the effect of residual variation. However, the author works under the strong assumption that at least one coefficient is the same in both groups. Therefore, I follow \cite{Long} and conduct tests of predicted probabilities -- as given in eq. \ref{eq:4} -- for comparing control and treatment groups. The advantage is that the predicted probabilities are unaffected by group differences in residual variation. Columns (1)-(3) of Table \ref{table5} report the predictions of average employment probabilities for the four different phases of child age for the control and treatment samples, their respective differences, and the $p$-values for the test in eq. \ref{eq:4}.\footnote{The average employment predictions show the counterfactual situation in which mothers do not have a second child, i.e., the predictions for the four child age phases are purely driven by a first child and lagged employment.}
\vspace{0,4cm}
\begin{center}
	[Insert Table \ref{table5} here.]
\end{center}
\vspace{0,4cm}
The results of the average employment predictions are in line with the expected labor market effects after childbirth: For both the control and treatment samples, the employment predictions are gradually increasing with the age of the first child. The parental benefit reform's direct policy effect on maternal employment is quite pronounced: mothers in the treatment sample display a 10 percentage points lower average employment rate (statistically significant) during the first 14 months after childbirth compared to mothers of the control sample. There are small differences in the average employment rates of the pre- and post-reform mothers in Phases II and IV. In Phase III, there is no statistically significant difference. Overall, for the post-reform sample, I find a small and statistically significant drop in the average employment probability over the entire sample period, driven mainly by the parental benefit reform's direct policy effect. 

One drawback of the baseline estimation from above is that it remains silent about the latent heterogeneity in maternal labor supply dynamics. However, it is precisely this latent heterogeneity that opens up a new perspective on the ongoing debate on the impact of childbirth on maternal employment. Therefore, I next want to identify the latent structures in the mothers' return-to-employment behavior after first birth by using the framework of the panel structure model in Section \ref{Model}. This approach allows the first child's age, along with the second child and lagged employment, to have heterogeneous effects across mothers, which are classified into groups with the C-Lasso technique. Such heterogeneous employment effects across mothers in childbirth response have useful implications for policymakers in designing efficient parental leave legislation.

\subsection{Results for heterogeneous employment effects}
\subsubsection*{Mixed Panel Structure model}
\begin{figure*}
		\centering
		\includegraphics[width=20cm]{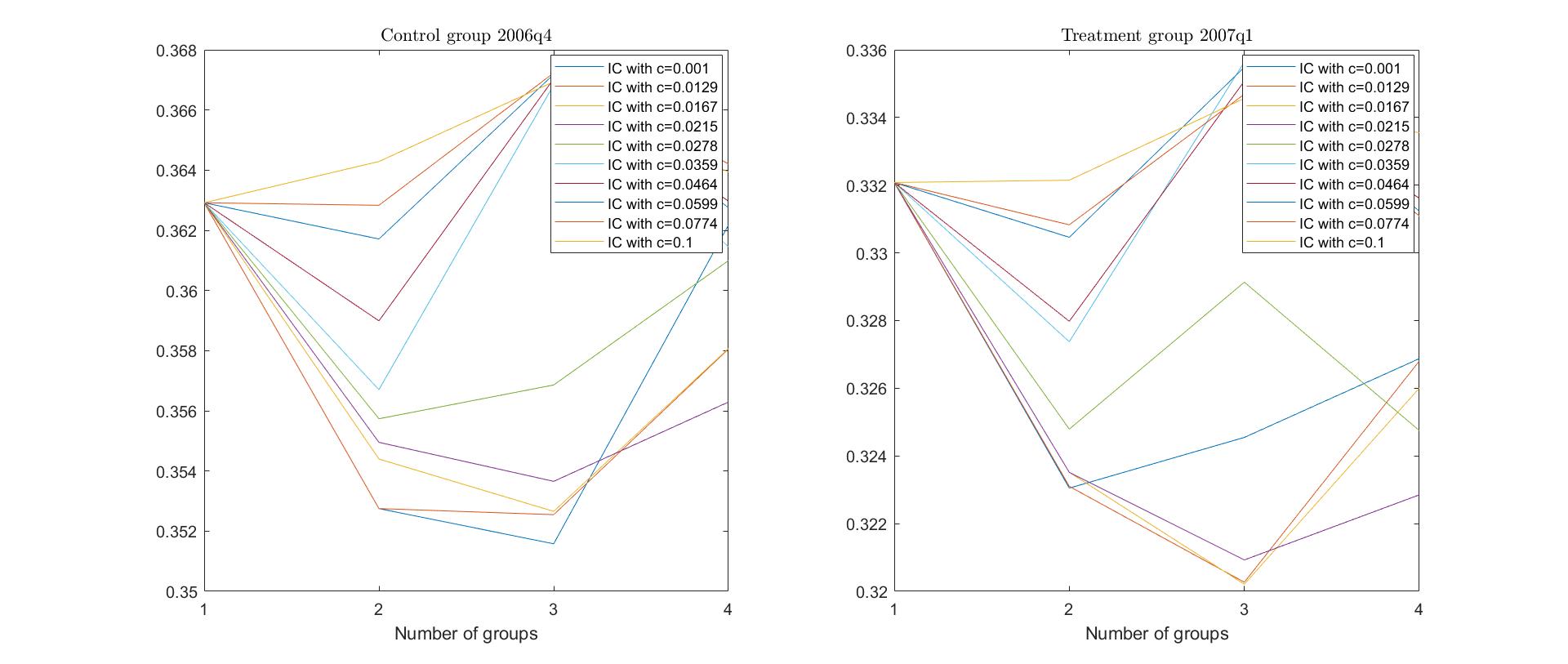}
		\caption{\small The horizontal axis depicts the number of groups and the vertical axis depicts the corresponding information criterion value.}
		\label{fig1}
	\end{figure*}

First, I use the C-Lasso approach suggested by \cite{Su16} to identify the latent group structure in eq. \ref{eq:2}. Regarding the C-Lasso tuning parameter, I choose $\lambda = c_{\lambda} \times s^2_{employ} \times T^{-1/3}$, where $s^2_{employ}$ is the sample variance of the employment indicator, $T$ is the number of time periods,\footnote{Throughout the estimations, I consider $T=73$.} and $c_{\lambda}$ is element of a 10 points geometrically increasing sequence $(0.01, \dots, 0.1)$. I allow for a maximum number of 4 groups. I calculate the information criterion value for each combination of the number of groups $K$ and the tuning parameter $c_{\lambda}$ according to eq. \ref{eq:24} of the methodology section. The number of groups and the tuning parameter are chosen by minimizing the information criterion from the set of all possible combinations. Fig. \ref{fig1} shows that the lowest point of the information criterion is obtained when $K=3$ and $c_{\lambda}=0.0129$ for the control group, and $K=3$ and $c_{\lambda}=0.01$ for the treatment group.  

Applying C-Lasso to the control ($2006m10-m12$) and treatment samples ($2007m1-m3$), I identify three latent groups for both samples. Columns (5)-(10) of Tables \ref{table9} and \ref{table4} report the post Lasso-PPL coefficient estimates and the corresponding marginal effects for each group. All post-Lasso-PPL estimates are bias-corrected by the half-panel jackknife, and the standard errors are again clustered at the individual level. Tables \ref{table9} and \ref{table4} suggest that the estimates for the slope coefficient on lagged employment are relatively stable across the three groups. Moreover, the estimates on the lagged employment variable are quite large in magnitude, indicating a strong persistence in labor supply decisions. In contrast, the estimates of the slope coefficients of the impact of child age on employment vary substantially across the three identified groups and even alter the signs. The $t$-statistics reported in Table \ref{table99} show that for Group 3, all post-Lasso estimates of the child age phases are statistically different from each other.  However, for Group 1 and Group 2, I cannot confirm that all coefficient estimates are statistically different from each other.

Before disentangling the reform effect, I report the group-specific marginal effects obtained by the C-Lasso.  The marginal effects derived from the group-specific post-Lasso-PPL estimates are all statistically significant. Starting with the interpretation of the fourth child age phase, I identify a quite heterogeneous pattern for the control and treatment samples: a group of mothers with a very low estimated employment probability (Group 1: 9.4\%/8.9\%), a group of mothers with a moderate estimated employment probability (Group 2: 33.9\%/31\%) and a group of mothers with a high estimated employment probability (Group 3: 59.4\%/58\%) when having a child aged between 60 and 72 months and is not employed previous month. My results for the control and treatment samples of Group 1 show that having a child aged between 1 and 14 months instead of a child aged between 60 and 72 months decreases, and a child aged between 15 and 24 months (/ between 25 and 59 months) increases maternal employment by on average 0.59/1.27, 6.2/5.5 and 4.5/8.7 percentage points holding constant all other factors. The corresponding estimates for Group 2 are -11.9/-18, -6.1/-5.4, and -6.1/-7. Finally, for the control and treatment samples of Group 3, I find that having a child aged between 1 and 14 (/ between 15 and 24 / between 25 and 59) months instead of a child aged between 60 and 72 months decreases maternal employment by on average 44.3/43, 29.3/24.3 and 11.5/8.9 percentage points holding constant all other factors. Overall, the marginal effects of the child age phases from the C-Lasso estimation suggest that mothers' identified groups show a highly heterogeneous dynamic in the return-to-employment behavior. I can identify three different cluster groups, which are sufficiently distinct, both in statistical terms and in terms of allowing for a meaningful interpretation.

Columns (4)-(12) of Table \ref{table5} report the group-specific predictions of the average employment probabilities for the four child age phases for the control and treatment samples, their corresponding differences $\Delta_{T-C}$ and the $p$-values of the test in square brackets underneath. In an attempt to interpret the cluster-groups based on the magnitude of the predicted employment probabilities and the evidence from previous literature, a clear pattern arises: Group 1 consists of mothers who are, on average, less employed than the mothers of the other two groups. Moreover, this cluster has a mobile dynamic of employment as it switches from less to more to the less average employment rate. Group 2 could consist of work-oriented mothers who decide immediately after benefit exhaustion to return to employment. Group 3 can be interpreted as a cluster of mothers who take an extended family break after childbearing, which goes far beyond the government-subsidized maternity leave period. However, on the other hand, most mothers of Group 3 have returned to employment until the age of six of the first child, i.e., most mothers return to employment during the 36 months of post-birth job protection. 

\begin{figure*}
	\centering
	\includegraphics[width=14cm]{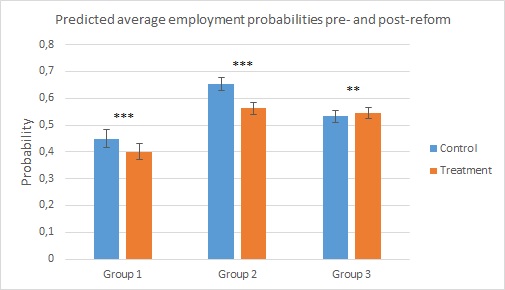}
	\caption{\small }
	\label{fig4}
\end{figure*}

Next, I test the group-specific differences of the four child age phases for statistical significance. I again conduct tests of predicted probabilities to compare the three identified groups' respective control and treatment samples. First, it is noticeable that the parental benefit policy affects the three groups differently. In fact, policy reform affects Group 2 the most. The direct and the indirect policy effects of the parental benefit reform on maternal employment are quite pronounced because I get evidence of significant average employment probability drops, ranging between 5-23 percentage points.\footnote{Here, I am careful to interpret these effects as causal, as this only works under strict identification assumptions.} For Group 1, I also find large negative direct and indirect policy effects, but on the other hand, I find no drop in predicted employment in the medium-run. This pattern has the consequence that I find an overall smaller drop in the average employment probability (4.8 p.p.) compared to the second group (9 p.p.). For Group 3, I even find a much smaller drop in average employment probability of 3.3 percentage points during Phase I and a pronounced increase in employment probabilities of 5.1 and 3.1 percentage points in Phases II and III. This leads to a statistically significant increase in average employment probability by 1.2 percentage points over the entire period (72 months after birth). Fig. \ref{fig4} shows the reform effect for the three identified groups over the entire period.

Finally, I discuss some descriptive statistics of the three groups that allow for further interesting insights from the C-Lasso group classification. Table \ref{table6} reports the number of mothers selected in each group. The C-Lasso method classifies over 50\% of the considered mothers in Group 3 for both the control and treatment samples. Only about 20\% of mothers are in the first group and almost 30\% in the second group. My finding confirms the evidence of previous literature (e.g., \cite{Schoenberg2014}) that most mothers make use of the generous government-provided partially paid leave in Germany to spend more time with their infants after childbirth.\footnote{In Germany, women are currently eligible for three years of partially paid and job-protected leave.} Maternal employment is lowered only in the short- and medium-run. Almost all mothers of Group 3 return back to employment with the exhaustion of government-provided job protection. However, there is still a significant proportion of mothers with a more atypical pattern of return to employment that has not been identified by the quasi-experimental literature. Indeed, 20\% of mothers decide to work much less on average than the other cluster-groups. On the other hand, 30\% of mothers have a permanently high average employment level after the first child age phase. 
\begin{table*}[h!]
	\begin{center}
		\caption{Number of observations of the heterogenous groups}\label{table6}
		\begin{tabular}{p{1.4cm} >{\raggedleft}p{1.5cm} >{\raggedleft}p{1.5cm} p{0.1cm} >{\raggedleft}p{1.2cm} >{\raggedleft}p{1.2cm} p{0.1cm} >{\raggedleft}p{1.2cm} >{\raggedleft}p{1.2cm} p{0.1cm} >{\raggedleft}p{1.3cm} p{1.3cm}<{\raggedleft} }  \hline \\[-1.0em]
			&  \multicolumn{11}{c}{Number of observations}  \\[0,05cm]
			& \multicolumn{2}{c}{Full sample}  & & \multicolumn{2}{c}{Group 1} & &\multicolumn{2}{c}{Group 2} & & \multicolumn{2}{c}{Group 3} \\\cline{2-3} \cline{5-6} \cline{8-9} \cline{11-12}\\[-1.0em]
			by & Control & Treatment & & Control & Treatment & & Control & Treatment & & Control & Treatment \\ \hline \\[-0.08em]
			\# obs. & 329 (100 \%) & 333 (100 \%) & & 65 \ (20 \%) & 68 \ (20 \%) & & 95 \ (29 \%) & 83 \ (25 \%) & & 169 (51 \%) & 182 (55 \%) \\ \hline \\[0,05cm] 
			\multicolumn{9}{l}{\small  }		
		\end{tabular}
	\end{center}
\end{table*}
\begin{table*}[h!]
	\begin{center}
		\caption{Descriptives of $2nd$ child}\label{table7}
		\begin{tabular}{p{2cm} >{\raggedleft}p{1.3cm} >{\raggedleft}p{1.3cm} p{0.1cm} >{\raggedleft}p{1.3cm} >{\raggedleft}p{1.3cm} p{0.1cm} >{\raggedleft}p{1.3cm} >{\raggedleft}p{1.3cm} p{0.1cm} >{\raggedleft}p{1.3cm} p{1.3cm}<{\raggedleft}}  \hline \\[-1.0em]
			& \multicolumn{11}{c}{Number of observations} \\[0.05cm]
			& \multicolumn{2}{c}{Total}		& & \multicolumn{2}{c}{Group 1} &  &\multicolumn{2}{c}{Group 2} & & \multicolumn{2}{c}{Group 3} \\\cline{2-3} \cline{5-6} \cline{8-9}  \cline{11-12} \\[-1.0em]
			by &Control & Treatment & & Control & Treatment & & Control & Treatment & & Control & Treatment  \\ \hline \\[-0.08em]
			child$_{\text{m1-m14}_{it}}$ & 11 & 12& & 1 &3 & & 2 & 0 & & 8 & 9 \\
			child$_{\text{m15-m24}_{it}}$ & 33	& 45 & & 7 & 11 & & 10 & 6 & & 16 & 28 \\
			child$_{\text{m25-m59}_{it}}$ & 110 & 121 & & 22 & 33 & & 43 & 45 & & 45 & 43\\
			child$_{\text{m60-m72}_{it}}$ & 18 & 21 & & 10 & 8 & & 6 & 8 & & 2 & 5 \\ \hline \\[-1.0em]
			$\Sigma $ & 172  (52 \%) & 199 (60 \%) & & 40 \ \ (62 \%) & 55 \ \ (81 \%)  & & 61 \ \ (64 \%) & 59 \ \ (71 \%) & & 71 \ \ (42 \%) & 85 \ \ (47 \%)   \\ \hline \\[-1.0em]
			\multicolumn{12}{l}{\small * The percentages reported here set the absolute numbers in relation to the number of observations}	\\
			\multicolumn{12}{l}{\small in each group.}	
		\end{tabular}
	\end{center}
\end{table*}

\cite{Fr16} have also identified cluster groups similar to those in this paper. The authors obtained five different clusters of mothers: a `high wage' cluster characterized by a quick return to employment, a cluster of `mobile careers' switching in and out of employment several times, an `extended family break' cluster, a `drop out of the labor force' cluster and a `part-time working' cluster. I cannot identify an `out of labor force' cluster similar to that of \cite{Fr16} because I exclude mothers from my sample who are never employed after their first birth.\footnote{Mothers without employment in the sample period are dropped from the data since they are associated with infinite ($\infty$ or $-\infty$) fixed effects in a FE probit model. cf. the Appendix for more details on this issue.} A `part-time' cluster can also not be identified within my framework because, unlike \cite{Fr16}, I do not consider a multinomial model. However, in contrast to \cite{Fr16}, I can determine the group classification and estimation of the coefficients simultaneously. The approach of \cite{Fr16}, on the other hand, is purely descriptive, i.e., they do not obtain coefficient estimates from their Bayesian classification method.

The estimation results of the average employment probability predictions confirm the  parental benefit reform's expected labor market effects in 2007. In line with the previous literature (e.g., \cite{Kluv18}, \cite{Kluve2013}), I find that the generous income replacement of the parental benefit reform generates a strong disincentive-to-work effect up to 14 months after childbirth. \cite{Kluv18} has shown that this strong disincentive-to-work effect is especially driven by those mothers who financially benefited from the reform. Their results show that the effects on the employment probability of mothers are positive and significant. Overall, I can identify three cluster groups with similar group-specific estimates for both the control and treatment samples. Thus, the parental benefit reform does not seem to change the composition of groups. Nevertheless, if the estimates within each group between control and treatment samples are compared, it seems that the parental benefit reform changes how the age of a child within each group influences maternal employment.  My results also confirm that the disincentive-to-work effect during Phase I is heterogeneous across groups and strongest for Group 1 and Group 2. However, for the third group, on the other hand, I find evidence of a significant increase in employment rates during Phase II and Phase III. \cite{Kluv18} also found evidence of pronounced beneficial medium run labor market effects of the parental benefit, especially for mothers from the upper tercentile of the income distribution. My findings suggest that the mothers considered here exhibit enormous diversities in both the return-to-employment behavior and the response to the parental benefit reform. In contrast to the recent literature (e.g., \cite{Kluv18}), the diversity identified here is generated by unobserved heterogeneity in the effects of child age on employment. A conclusion on the main drivers, similar to an analysis based on observable heterogeneity, is therefore not possible.    

If I pool groups 1-3 together and estimate the model in \ref{eq:2}, I find that the pooled estimate can be interpreted as a weighted average of the three group-specific estimates. However, this weighted average remains silent about the latent heterogeneous pattern in the effects of child age on maternal employment provided by the C-Lasso method. In particular, estimating average effects hide important and policy-relevant differences in the employment trajectories. As we have already seen, the reform generates strong negative effects for Group 2, whereas it encourages the mothers of Group 3 to take up employment immediately after benefit exhaustion. My findings confirm the importance of considering latent structures in the data, which may help design an efficient parental leave policy.
\subsubsection*{Robustness checks}
In this subsection, I present two further specifications exploring the robustness of the main estimates of the heterogeneous employment effects on the functional form and panel structure model chosen. First, specification A runs C-Lasso for a nonlinear fixed effects model without controlling for the dynamic component and the $2nd$ child variable. In a second robustness check, I estimate the mixed panel structure model from above without controlling for the $2nd$ child (specification B).

I check for the robustness of my findings using alternative identifying assumptions because applied researchers using panel data are often confronted with the challenge of choosing between dynamic and fixed-effects models. They usually avoid using both identifications in one model like in my approach since the conditions for consistently estimating causal effects are much more demanding than those required with lagged dependent variables or fixed effects alone.\footnote{The challenge of consistent estimation in a dynamic nonlinear fixed-effect model can be easily seen when taking the first difference to eliminate the fixed effects and then estimating the differenced model with OLS. The parameter estimates are inconsistent as the differenced residual is correlated by construction with the lagged dependent variable, a problem first noted by \cite{Nickell1981}.} Even though I apply the split-panel jackknife method suggested by \cite{Dhaen15}, which is explicitly designed to handle dynamics in the data and to reduce the bias induced by the incidental parameter problem (\cite{Ney48}), I would like to find out if I can find broadly similar results using a plausible alternative specification.

\begin{center}
	\begin{figure*}
		\includegraphics[width=20cm]{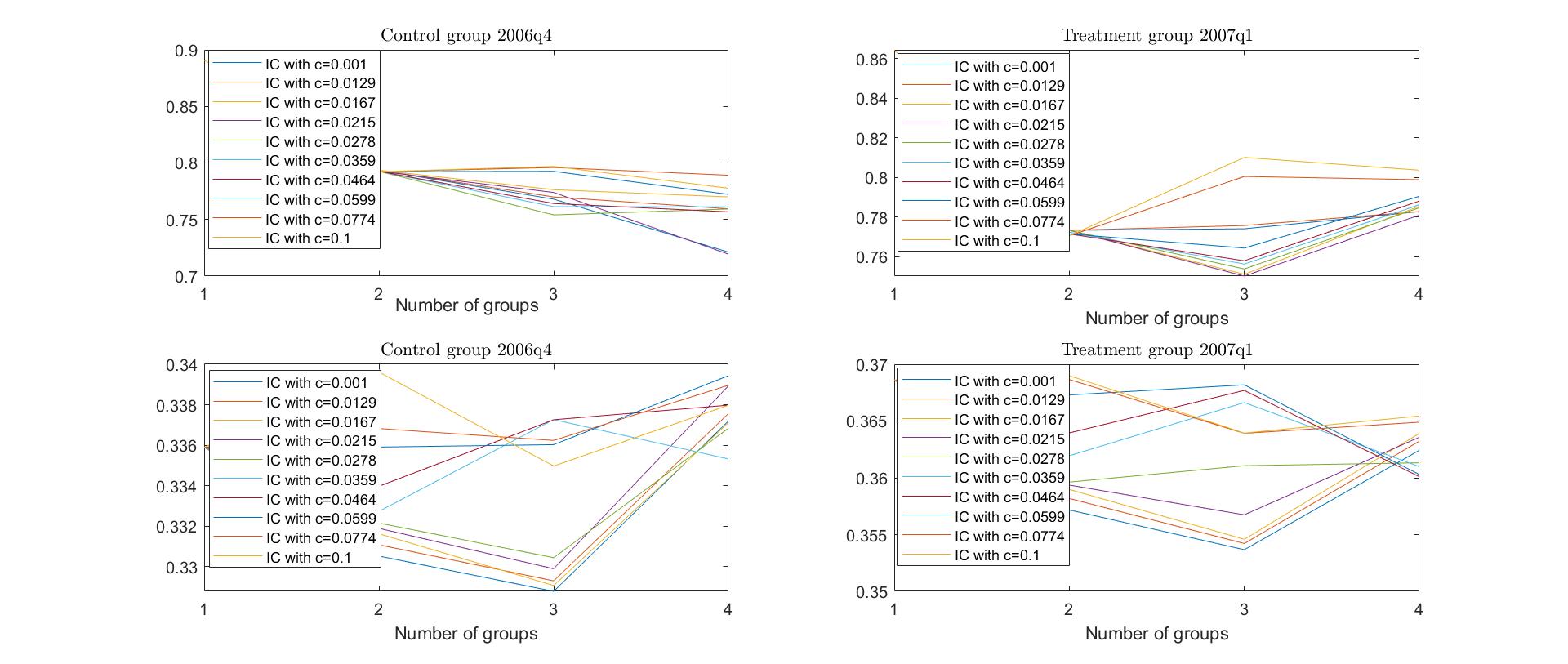}
		\caption{\small The horizontal axis depicts the number of groups and the vertical axis depicts the corresponding information criterion value. The upper panels show the information criterion values of the first robustness check and the lower panels show the information criterion values of the second robstness check.}
		\label{fig2}
	\end{figure*}
\end{center}

For the first robustness check, I adapt the panel structure model in \ref{eq:3} by dropping $employ_{i,t-1}$ and $child_{2nd_{it}}$ from the model. Then I again use the C-Lasso to identify latent group structures in the model. By applying C-Lasso to the treatment sample, I still find three latent groups for the adapted model in \ref{eq:3}. For the control sample, even four latent groups are identified.\footnote{For reasons of the empirical approach's assumptions, however, only results for $K=3$ are reported.} Figure \ref{fig2} plots the information criterion function where the vertical and horizontal axes mark the information criterion values and the number of groups, respectively. The lowest point of the information criterion is achieved when $K=4$ and $c_{\lambda}=0.0215$ for the control sample, and $K=3$ and $c_{\lambda}=0.0215$ for the treatment sample.
\vspace{0,4cm}
\begin{center}
	[Insert Figure \ref{fig2} here.]
\end{center}
\vspace{0,4cm}

The upper panel of Table \ref{table13} reports the estimation results of the post-Lasso-PPL coefficients for each group in model \ref{eq:3} and the pooled version of this model. 
\vspace{0,4cm}
\begin{center}
	[Insert Table \ref{table13}, Table \ref{table14} and Table \ref{table15} here.]
\end{center}
\vspace{0,4cm}
The corresponding marginal effects and predicted average employment probabilities are found in the upper panels of Tables \ref{table14} and \ref{table15}. The tables show that the identified groups are sufficiently distinct in terms of estimated child age effects and similar to those obtained by the main mixed panel structure model reported in the previous subsection. The estimation results suggest that the group structure is stable when all child age parameters $\gamma_i, \delta_i$, and $ \eta_i$ are again allowed to be heterogeneous across mothers, and the dynamic component and the $2nd$ child variable are omitted from the model. In particular, I again identify a group of mothers (Group 1) who are faced with overall lower predicted employment probabilities than the other two groups, a group of mothers (Group 2) who timely return to employment after childbirth, and a group of mothers (Group 3) taking an extended family break before returning to employment. The only difference I can state is that I get lower predicted average employment probabilities for Group 1 and larger negative child age effects, obviously driven by the two omitted variables. The robustness check confirms that there should be no major concern regarding the specification of a dynamic nonlinear fixed effects model used in this paper. 

The estimation results of specification B are reported in the lower panels of Table \ref{table13}, Table \ref{table14} and Table \ref{table15}. As shown in Figure \ref{fig2}, the information criterion again selects three groups for the control and treatment samples. The three identified groups are again fairly similar to those obtained in the main estimation results. This means that there should be no major concern that subsequent fertility decisions (here for a second child) could be the main drivers of the C-Lasso group classification. 

The results obtained by the mixed panel structure model and the two robustness checks are in stark contrast to the conclusions of \cite{Kluv18}. The authors argue that the new parental benefit reform defines an institutionalized point in time at which mothers go back to work. In Germany, unpaid parental leave entitlements cover the time period up to three years after childbirth \cite{Schoenberg2014}. According to \cite{Kluv18}, the exact time point to return to the job was an individual decision under the previous regulation. However, after the policy change, many working mothers return to the job at the end of the maximum benefit receipt. The authors found evidence that the benefit reform induced a strong homogenization of time spent on paid parental benefit. My results, however, suggest that the parental benefit reform does not change the composition of groups but only changes how the age of a child within each group influences maternal employment. My findings reveal - pre- and post-reform - three highly heterogeneous groups of maternal return-to-employment behavior. It is noticeable that the largest group of mothers (Group 3) takes an extended family break after childbirth, which goes far beyond the government-provided paid leave.

\section{Conclusion}\label{Conclusion}
In this paper, I identify latent structures in maternal employment using the C-Lasso methodology, a new data-driven classification approach proposed by \cite{Su16}. In particular, I assess how the introduction of the generous German parental benefit reform in 2007 affects mothers' different clusters. By exploiting an identification strategy that combines the sharp RD approach and hypothesis testing of predicted employment probabilities, I can identify the impacts of the policy reform on employment within the different groups. 

Using a new data basis of German custom shaped administrative employment records, the C-Lasso mechanism identifies three different cluster groups pre- and post-reform. My results not only reveal marked heterogeneity in the effect of child age on employment across mothers but also that the German policy change affects the employment patterns of the three cluster groups differently. Especially noteworthy is that under the new policy, the largest group of mothers is faced with comparable small negative employment changes during the benefit receipt and even exhibits an increase in average employment rates after exhaustion of government-provided transfers. But the findings also reveal that under the new policy, one group of mothers suffers from large and long-lasting negative employment changes after childbirth. The data-determined grouping results provide some new insights about potential impacts of child age on maternal employment in general, and more specifically, how the reform affects these heterogeneous groups. My results suggest that the identified latent heterogeneity in maternal employment should be taken into account by policymakers to design efficient parental leave legislation. My evidence also has important policy implications beyond the German case, as latent group structures in maternal employment are likely to be found in other countries as well.

Finally, I note that employment losses of women associated with first childbirth are still quite pronounced in Germany and that for Group 1 and Group 2, the parental leave reform is not able to compensate for the strong disincentive-to-work effect in Phase I by providing a sufficiently strong incentive-to-work effect after benefit exhaustion. Therefore, a policy should particularly pay attention to these two groups, which involve around 50\% of mothers.

\newpage
\bibliographystyle{plainnat}
\bibliography{Literatur} 
\renewcommand\refname{Literaturverzeichnis} 

\section{Appendix}		
The appendix consists of two main parts. The first part provides references to the code material used in this study. The second part contains the main tables of my estimations.	         
\subsection*{Code material}
\cite{Su16} provide documented code material for their simulations and empirical data applications online. Their coding material used for implementing the C-Lasso within panel structure models is available for the Matlab software. I have used their code material, rewritten the code for my purposes, and extended it to implement the C-Lasso for mixed-panel structure models, which were also estimated in my study. \\[5cm]

\subsection*{Tables}
This subsection lists all the tables that contain the results of the main estimations.
\begin{sidewaystable*}[h!]
	\begin{center}
		\caption{Marginal Effects of Employment Effects}\label{table3}
		\begin{tabular}{p{3cm} >{\raggedleft}p{3cm} >{\raggedleft}p{3cm} p{0.2cm} >{\raggedleft}p{7.2cm} >{\raggedleft}p{3cm} p{3cm}<{\raggedleft}}  \hline \\[-1.0em]
			& \multicolumn{2}{c}{LPM FE} & & \multicolumn{3}{c}{Testing linear combinations}  \\[0,05cm]   \\[-1.0em]		
			& \multicolumn{2}{c}{Full sample} & & Null hypothesis & $p$-Value  & $p$-Value  \\[0,05cm] \cline{2-3} \cline{5-7} \\[-1.0em]	
			&	Control mean & Treatment mean & & & Control sample & Treatment sample \\
			& (1) & (2) & & (3)  & (4) & (5) \\\hline \\[-1.0em]
			employ$_{it-1}$&  0.8327 (0.0057)     & 0.8073 (0.0068)    & & $\text{child}_{\text{m1-m14}_{it}}-$ child$_{\text{m15-m24}_{it}}=0$& 0.0000 & 0.0000	\\[0,1cm]
			child$_{\text{2nd}_{it}}$&  -0.0367 (0.0059)     & -0.0509 (0.0068)   & & child$_{\text{m1-m14}_{it}}-$ child$_{\text{m25-m59}_{it}}=0$& 0.0000  & 0.0000  \\[0,1cm]
			child$_{\text{m1-m14}_{it}}$& -0.0838 (0.0064)    &  -0.1144 (0.0076)   & & child$_{\text{m15-m24}_{it}}-$ child$_{\text{m25-m59}_{it}}=0$& 0.0000  & 0.0000 \\[0,1cm]
			child$_{\text{m15-m24}_{it}}$& -0.0499 (0.0056)    &  -0.0550 (0.0057)   & &child$_{{\text{m1-m14}_{it}}_T}-$ child$_{{\text{m1-m14}_{it}}_C}=0$  &  \multicolumn{2}{c}{0.0020}    \\
			child$_{\text{m25-m59}_{it}}$& -0.0225 (0.0035)     & -0.0220 (0.0042)    &  &child$_{{\text{m15-m24}_{it}}_T}-$ child$_{{\text{m15-m24}_{it}}_C}=0$  & \multicolumn{2}{c}{0.5275}   \\[0,1cm]
			 &  &  & & child$_{{\text{m25-m59}_{it}}_T}-$ child$_{{\text{m25-m59}_{it}}_C}=0$ & \multicolumn{2}{c}{0.9275}        \\ \hline \\[-1.0em]	    
			\multicolumn{7}{l}{\small \textit{Notes}: LPM = Linear Probability Model, FE = Fixed Effects.}	\\
			\multicolumn{7}{l}{\small  Standard errors are provided in (parentheses) and are clustered at the individual level.} \\	
			\multicolumn{7}{l}{\small Estimation with Stata. I use the lincom command to test whether linear combinations of parameter estimates within the same sample are statistically  } \\
			\multicolumn{7}{l}{\small different from each other.  I use the \texttt{suest} command to test the equality of two parameter estimates in the independent treatment and control samples. } \\
			\multicolumn{7}{l}{\small I account for individual fixed effects by including a separate dummy for each sample. Estimation without intercept, because all fixed effects are included.} 
		\end{tabular}
	\end{center}
\end{sidewaystable*}

\clearpage
\begin{sidewaystable*}[h!]
	\begin{center}
		\caption{Marginal Effects of Employment Effects}\label{table33}
		\begin{tabular}{p{3cm} >{\raggedleft}p{3cm} >{\raggedleft}p{3cm} p{0.2cm} >{\raggedleft}p{7.2cm} >{\raggedleft}p{3cm} p{3cm}<{\raggedleft}}  \hline \\[-1.0em]
			& \multicolumn{2}{c}{LPM} & & \multicolumn{3}{c}{Testing linear combinations}  \\[0,05cm]   \\[-1.0em]		
			& \multicolumn{2}{c}{Full sample} & & Null hypothesis & $p$-Value  & $p$-Value  \\[0,05cm] \cline{2-3} \cline{5-7} \\[-1.0em]	
			&	Control mean & Treatment mean & & & Control sample & Treatment sample \\ 
				& (1) & (2) & & (3)  & (4) & (5) \\\hline \\[-1.0em]
			employ$_{it-1}$&  0.9105 (0.0038)  &  0.8858 (0.0053)  & & child$_{\text{m1-m14}_{it}}-$ child$_{\text{m15-m24}_{it}}=0$& 0.0000  	&  0.0000 	\\[0,1cm]
			child$_{\text{2nd}_{it}}$& -0.0191 (0.0038)  & -0.0329 (0.0048)  & & child$_{\text{m1-m14}_{it}}-$ child$_{\text{m25-m59}_{it}}=0$&  0.0000	& 0.0000  \\[0,1cm]
			child$_{\text{m1-m14}_{it}}$&  -0.0448 (0.0048)  & -0.0693 (0.0062)     & & child$_{\text{m15-m24}_{it}}-$ child$_{\text{m25-m59}_{it}}=0$&  0.0000  & 0.0000  \\[0,1cm]
			child$_{\text{m15-m24}_{it}}$& -0.0265 (0.0041)  &  -0.0312 (0.0046)   & &child$_{{\text{m1-m14}_{it}}_T}-$ child$_{{\text{m1-m14}_{it}}_C}=0$  &  \multicolumn{2}{c}{0.0017}    \\[0,1cm]
			child$_{\text{m25-m59}_{it}}$& -0.0115 (0.0027)    & -0.0137 (0.0032)  &  &child$_{{\text{m15-m24}_{it}}_T}-$ child$_{{\text{m15-m24}_{it}}_C}=0$  & \multicolumn{2}{c}{0.4419}   \\[0,1cm]
			Intercept & 0.0672 (0.0047) &  0.0880 (0.0063)  & & child$_{{\text{m25-m59}_{it}}_T}-$ child$_{{\text{m25-m59}_{it}}_C}=0$ & \multicolumn{2}{c}{0.6134}        \\[0,1cm]           & & & &  cons$_T-$cons$_C=0$ & \multicolumn{2}{c}{0.0082}  \\[0,1cm] \hline \\[-1.0em]			
			\multicolumn{7}{l}{\small \textit{Notes}: LPM = Linear Probability Model.}	\\
			\multicolumn{7}{l}{\small  Standard errors are provided in (parentheses) and are clustered at the individual level.} \\	
			\multicolumn{7}{l}{\small Estimation with Stata. I use the \texttt{lincom} command to test whether linear combinations of parameter estimates within the same sample are statistically  } \\
			\multicolumn{7}{l}{\small different from each other. I use the \texttt{suest} command to test the equality of two parameter estimates in the independent treatment and control samples.} 
		\end{tabular}
	\end{center}
\end{sidewaystable*}

\clearpage
\begin{sidewaystable*}[h!]
	\begin{center}
		\caption{Effects on Employment Rates}\label{table9}
		\begin{tabular}{p{2.3cm}  >{\raggedleft}p{1.5cm}  >{\raggedleft}p{1.5cm} p{0.2cm} >{\raggedleft}p{1.5cm} >{\raggedleft}p{1.5cm} p{0.2cm} >{\raggedleft}p{1.5cm} >{\raggedleft}p{1.5cm} p{0.2cm} >{\raggedleft}p{1.5cm} >{\raggedleft}p{1.5cm} p{0.2cm} >{\raggedleft}p{1.5cm} p{1.5cm}<{\raggedleft}}  \hline \\[-1.0em]
			&\multicolumn{2}{c}{Probit FE} &	&  \multicolumn{11}{c}{Post Lasso-PPL }  \\[0,05cm] \cline{2-3} \cline{5-15} \\[-1.0em]
			& \multicolumn{2}{c}{Full sample} & & \multicolumn{2}{c}{Full sample} &	& \multicolumn{2}{c}{Group 1} &  &\multicolumn{2}{c}{Group 2} &  & \multicolumn{2}{c}{Group 3} \\  \cline{2-3} \cline{5-6} \cline{8-9} \cline{11-12} \cline{14-15} \\[-1.0em]
			&	Control mean & Treatment mean & &Control mean & Treatment mean & &Control mean & Treatment mean & & Control mean & Treatment mean & & Control mean & Treatment mean \\
			& (1) & (2) & & (3)  & (4) & & (5) & (6) & & (7) & (8) & & (9) & (10) \\ \hline \\[-1.0em]
			employ$_{it-1}$& 2.9476 (0.0374) & 2.7611 (0.0389) & &3.2094 (0.0366)& 3.0192 (0.0383)  & & 2.8172 (0.0734) & 2.5531 (0.0801) & & 2.8833 (0.0637) & 2.6880 (0.0672) & &  3.0922 (0.0561) & 3.0225 (0.0571)                  \\
			child$_{\text{2nd}_{it}}$& -0.5253 (0.0610) & -0.6059 (0.0549) & &-0.2159 (0.0601)& -0.2835 (0.0541)  & & -0.4693 (0.1233) & -0.5731 (0.0972) & & -0.2081 (0.0948) & -0.0123 (0.0926) & & -0.3406 (0.1110) & -0.1742 (0.0944)    \\
			child$_{\text{m1-m14}_{it}}$& -1.1139 (0.0732) & -1.3068 (0.0687) & &-1.0069 (0.0715)& -1.1994 (0.0671)  & & -0.0396 (0.1231) & -0.0824 (0.1149) & & -0.7882 (0.1253) & -0.9967 (0.1196) & & -2.8898 (0.1543) & -2.6778 (0.1341)  \\
			child$_{\text{m15-m24}_{it}}$& -0.7193 (0.0685) & -0.7195 (0.0629) & &-0.7259 (0.0670)& -0.7152 (0.0617) & & 0.3984 (0.1183) & 0.3381 (0.1167) & & -0.4544 (0.1183) & -0.3591 (0.1156) & & -2.2118 (0.1297) & -1.8772 (0.1127)     \\
			child$_{\text{m25-m59}_{it}}$& -0.3365 (0.0490) & -0.3100 (0.0463) & &-0.3038 (0.0480)& -0.2904 (0.0456) & & 0.2942 (0.0884) & 0.5221 (0.0856) & & -0.4473 (0.0894) & -0.4602 (0.0854) & & -1.1117 (0.0953) & -0.8484 (0.0910)\\ \hline \\[-1.0em]
			\multicolumn{15}{l}{\small \textit{Notes}: FE = Fixed Effects, PPL= Penalized Profile Likelihood. All Post Lasso-PPL estimates are bias corrected by half-panel jackknife.}	\\
			\multicolumn{15}{l}{\small  Standard errors are provided in (parentheses) and are clustered at the individual level. Estimation without intercept, because all fixed effects are included.}	\\
		\end{tabular}
	\end{center}
\end{sidewaystable*}

\clearpage  
\begin{sidewaystable*}[h!]
	\begin{center}
		\caption{Supplements to Table \ref{table9}}\label{table99}
		\begin{tabular}{p{6.2cm}  >{\raggedleft}p{3cm} >{\raggedleft}p{2cm} >{\raggedleft}p{2cm} p{2cm}<{\raggedleft}}  \hline \\[-1.0em]
			& \multicolumn{4}{c}{Testing linear combinations ($t$-statistic)} \\[0,05cm]  	
			&  \multicolumn{4}{c}{Control sample}   \\[0,05cm] 
			& Full sample & Group 1 & Group 2 & Group 3 \\[0,05cm]
			& (1)   & (2)  & (3) & (4)  \\  \hline  \\[-1.0em]	
			child$_{\text{m1-m14}_{it}}-$ child$_{\text{m15-m24}_{it}}$& -2.73 &  -2.15 &  -2.06 & -3.13 \\[0,1cm]
			child$_{\text{m1-m14}_{it}}-$ child$_{\text{m25-m59}_{it}}$& -3.30 & -1.70 & -1.08 & -8.12 \\[0,1cm]
			child$_{\text{m15-m24}_{it}}-$ child$_{\text{m25-m59}_{it}}$& -2.05 & -0.13 &  0.55 & -5.74 \\ \hline \\[-1.0em]
			&  \multicolumn{4}{c}{Treatment sample}   \\[0,05cm] 
			& Full sample & Group 1 & Group 2 & Group 3 \\[0,05cm] 
			& (1)   & (2)  & (3) & (4)    \\ \hline  \\[-1.0em]	
			child$_{\text{m1-m14}_{it}}-$ child$_{\text{m15-m24}_{it}}$& -4.36 & -2.04 & -4.13 & -4.02 \\[0,1cm]
			child$_{\text{m1-m14}_{it}}-$ child$_{\text{m25-m59}_{it}}$& -4.50 & -2.66 & -2.15 & -8.32 \\[0,1cm]
			child$_{\text{m15-m24}_{it}}-$ child$_{\text{m25-m59}_{it}}$& -2.34 & -1.42 &  0.57 & -5.37 \\ \hline \\[-1.0em]				
		\end{tabular}
	\end{center}
\end{sidewaystable*}

\clearpage
\begin{sidewaystable*}[h!]
	\begin{center}
		\caption{Marginal Effects of Employment Effects}\label{table4}
		\begin{tabular}{p{2.3cm} >{\raggedleft}p{1.5cm} >{\raggedleft}p{1.5cm} p{0.2cm} >{\raggedleft}p{1.5cm} >{\raggedleft}p{1.5cm} p{0.2cm}  >{\raggedleft}p{1.5cm}  >{\raggedleft}p{1.5cm} p{0.2cm} >{\raggedleft}p{1.5cm} >{\raggedleft}p{1.5cm} p{0.2cm} >{\raggedleft}p{1.5cm} p{1.5cm}<{\raggedleft}}  \hline \\[-1.0em]
			& \multicolumn{2}{c}{LPM FE} & &  \multicolumn{11}{c}{Post Lasso-PPL }  \\[0,05cm] \cline{2-3} \cline{5-15 } \\[-1.0em]
			& \multicolumn{2}{c}{Full sample} & &  \multicolumn{2}{c}{Full sample} & &\multicolumn{2}{c}{Group 1} &  &\multicolumn{2}{c}{Group 2} &  & \multicolumn{2}{c}{Group 3} \\  \cline{2-3} \cline{5-6} \cline{8-9} \cline{11-12} \cline{14-15} \\[-1.0em]
			&	Control mean & Treatment mean &	&	Control mean & Treatment mean &	&	Control mean & Treatment mean & & Control mean & Treatment mean & & Control mean & Treatment mean \\
			& (1) & (2) & & (3)  & (4) & & (5) & (6) & & (7) & (8) & & (9) & (10) \\ \hline \\[-1.0em]
			employ$_{it-1}$& 0.8327 (0.0032)  & 0.8073 (0.0035)  & & 0.8032 (0.0051) & 0.7810 (0.0040) & & 0.7744 (0.0074) & 0.7168 (0.0084) & & 0.7563 (0.0101) & 0.7497 (0.0075) & & 0.6329 (0.0094) & 0.6214 (0.0085)        \\
			child$_{\text{2nd}_{it}}$& -0.0367 (0.0042) & -0.0509 (0.0045) & & -0.0263 (0.0003) & -0.0368 (0.0003) & & -0.0722 (0.0011) & -0.0939 (0.0016) & & -0.0286 (0.0004) & -0.0019 (0.0000) & & -0.0377 (0.0005) & -0.0198 (0.0003) \\
			child$_{\text{m1-m14}_{it}}$& -0.0838 (0.0045)  & -0.1144 (0.0051)  & & -0.1381 (0.0012) & -0.1809 (0.0009) & & -0.0059 (0.0001) & -0.0127 (0.0002) & & -0.1194 (0.0016) & -0.1797 (0.0012) & & -0.4430 (0.0030) & -0.4297 (0.0024) \\
			child$_{\text{m15-m24}_{it}}$& -0.0499 (0.0046) & -0.0550 (0.0050) & & -0.0883 (0.0009) & -0.0903 (0.0007) & & 0.0621 (0.0012) & 0.0545 (0.0012) & & -0.0614 (0.0009) & -0.0541 (0.0005) & & -0.2928 (0.0027) & -0.2434 (0.0025) \\
			child$_{\text{m25-m59}_{it}}$& -0.0225 (0.0032)  & -0.0220 (0.0035) & & -0.0367 (0.0004) & -0.0365 (0.0004) &  & 0.0454 (0.0007) & 0.0874 (0.0015) & & -0.0611 (0.0008) & -0.0698 (0.0007) & & -0.1154 (0.0010) & -0.0892 (0.0009) \\
			Intercept & 0.1226 (0.0042)  & 0.1430 (0.0047) & & 0.2638 (0.0091) & 0.2876 (0.0086) & & 0.0944 (0.0107) & 0.0892 (0.0087) & & 0.3388 (0.0173) & 0.3099 (0.0148) & & 0.5939 (0.0208) & 0.5804 (0.0201)    \\ \hline \\[-1.0em]
			\multicolumn{15}{l}{\small \textit{Notes}: LPM = Linear Probability Model, FE = Fixed Effects, PPL= Penalized Profile Likelihood. }	\\
			\multicolumn{15}{l}{\small Standard errors are provided in (parentheses), and are clustered at the individual level for the LPM FE (columns (1) and (2)).} \\
			\multicolumn{15}{l}{\small All standard errors and average effects derived from the Post Lasso-PPL estimates are build on a plug-in estimator.}	\\
		    \multicolumn{15}{l}{\small See the Appendix for further information on the plug-in principle used here.}	\\
			\multicolumn{15}{l}{\small Intercept is calculated ex-post as the average over all individual fixed effects belonging to the different groups.}	
		\end{tabular}
	\end{center}
\end{sidewaystable*}

\clearpage      
\begin{sidewaystable*}[h!]
	\begin{center}
		\caption{Prediction of Average Employment Probability}\label{table5}
		\begin{tabular}{p{2.3cm} >{\raggedleft}p{1.4cm}  >{\raggedleft}p{1.4cm} >{\raggedleft}p{1.4cm} p{0.1cm}   >{\raggedleft}p{1.4cm}  >{\raggedleft}p{1.4cm} >{\raggedleft}p{1.4cm} p{0.1cm} >{\raggedleft}p{1.4cm} >{\raggedleft}p{1.4cm} >{\raggedleft}p{1.4cm} p{0.1cm} >{\raggedleft}p{1.4cm} >{\raggedleft}p{1.4cm} p{1.4cm}<{\raggedleft}}  \hline \\[-1.0em]
			&  \multicolumn{15}{c}{Employment Probability}  \\[0,05cm]
			& \multicolumn{3}{c}{Full sample} & & \multicolumn{3}{c}{Group 1} &  &\multicolumn{3}{c}{Group 2} &  & \multicolumn{3}{c}{Group 3} \\  \cline{2-4} \cline{6-8} \cline{10-12} \cline{14-16} \\[-1.0em]
			&	Control mean & Treatment mean & $\Delta_{\text{T-C}}$ in p.p. & &	Control mean & Treatment mean & $\Delta_{\text{T-C}}$ in p.p. & & Control mean & Treatment mean & $\Delta_{\text{T-C}}$ in p.p.  & & Control mean & Treatment mean & $\Delta_{\text{T-C}}$ in p.p.  \\
			& (1) & (2) & (3) & & (4) & (5) & (6) & & (7) & (8) & (9) & & (10) & (11) & (12) \\ \hline \\[-1.0em]
			child$_{\text{m1-m14}_{it}}$& 0.2704 (0.0162) & 0.1637 (0.0105) & -10.67 [0.0000] & & 0.3156 (0.0358) & 0.2184 (0.0247) & -9.72 [0.0000] & & 0.4444 (0.0308) & 0.2131 (0.0210) & -23.13 [0.0000] & & 0.1437 (0.0194)  & 0.1108 (0.0137) & -3.29 [0.0000] \\
			child$_{\text{m15-m24}_{it}}$& 0.4880 (0.0234) & 0.4578 (0.0223) &-3.02 [0.0071] & & 0.5681 (0.0497) & 0.4509 (0.0471) & -11.72 [0.0000] & & 0.7613 (0.0337) & 0.6869 (0.0385) & -7.44 [0.0000] & & 0.2993 (0.0312)  &  0.3505 (0.0302) & 5.12 [0.0006]\\
			child$_{\text{m25-m59}_{it}}$& 0.6219 (0.0180) & 0.6260 (0.0178) & 0.41 [0.4692] & & 0.5351 (0.0401) & 0.5444 (0.0389) & 0.93 [0.4589] & & 0.6516 (0.0293) & 0.6065 (0.0297) & -4.51 [0.0000] & & 0.6316 (0.0273)  & 0.6624 (0.0261) & 3.08 [0.0000] \\
			child$_{\text{m60-m72}_{it}}$& 0.7497 (0.0190) & 0.7132 (0.0193)  & -3.65 [0.0000]  & & 0.2805 (0.0388) & 0.1869 (0.0279) & -9.36 [0.0000] & & 0.8108 (0.0266) & 0.7486 (0.0300) & -6.22 [0.0000] & & 0.8905 (0.0193)& 0.8731 (0.0199) & -1.74 [0.0414] \\ \hline \\[-1.0em]
			Ave. emp. &  0.5541 (0.0149) & 0.5235 (0.0140)  & -3.06 [0.0000] & & 0.4492 (0.0333) & 0.4009 (0.0301) & -4.83 [0.0000] & & 0.6524 (0.0241) & 0.5620 (0.0227)& -9.04 [0.0000] &    &  0.5319 (0.0219) & 0.5439 (0.0199)  &  1.2 [0.0402]                                                                            \\ \hline  \\[-1.0em]
			\multicolumn{16}{l}{\small \textit{Notes}: Average Employment Probability is calculated as $1/(N[\hat{\mathcal{G}}^k] T[p^j]) \ G(\boldsymbol{x}_{i[\hat{\mathcal{G}}^k]t[p^j]}^{\top} \boldsymbol{\hat{\hat{\alpha}}}_{\hat{\mathcal{G}}^k}+\hat{\mu}_i[\hat{\mathcal{G}}^k] $), where $G(\cdot)$ denotes the CDF of the standard normal,} 	\\
			\multicolumn{16}{l}{\small $p^j, \ j=1,\dots, 4$ denote the four different phases of childage, $[\cdot]$ denotes the restiction to a subspace, and the remainder is defined as above.}	\\
			\multicolumn{16}{l}{\small Standard errors are given in (parentheses) and are build on a plug-in estimator.}  \\
			\multicolumn{16}{l}{\small The $p$-values for the test in eq. \ref{eq:4} are reported in [square brackets] under the estimated differences of the average employment predictions.}	
		\end{tabular}
	\end{center}
\end{sidewaystable*} 

\clearpage 
\begin{sidewaystable*}[h!]
	\begin{center}
		\caption{Effects on Employment Rates}\label{table13}
		\begin{tabular}{p{2.3cm}  >{\raggedleft}p{1.5cm}  >{\raggedleft}p{1.5cm} p{0.2cm} >{\raggedleft}p{1.5cm} >{\raggedleft}p{1.5cm} p{0.2cm} >{\raggedleft}p{1.5cm} >{\raggedleft}p{1.5cm} p{0.2cm} >{\raggedleft}p{1.5cm} >{\raggedleft}p{1.5cm} p{0.2cm} >{\raggedleft}p{1.5cm} p{1.5cm}<{\raggedleft}}  \hline \\[-1.0em]
			&\multicolumn{2}{c}{Probit FE} &	&  \multicolumn{11}{c}{Post Lasso-PPL }  \\[0,05cm] \cline{2-3} \cline{5-15} \\[-1.0em]
			& \multicolumn{2}{c}{Full sample} & & \multicolumn{2}{c}{Full sample} &	& \multicolumn{2}{c}{Group 1} &  &\multicolumn{2}{c}{Group 2} &  & \multicolumn{2}{c}{Group 3} \\  \cline{2-3} \cline{5-6} \cline{8-9} \cline{11-12} \cline{14-15} \\[-1.0em]
			&	Control mean & Treatment mean & &Control mean & Treatment mean & &Control mean & Treatment mean & & Control mean & Treatment mean & & Control mean & Treatment mean \\ 
			& (1) & (2) & & (3)  & (4) & & (5) & (6) & & (7) & (8) & & (9) & (10) \\ \hline \\[-1.0em]
			child$_{\text{m1-m14}_{it}}$& -1.8453 (0.0969) & -2.1090 (0.0973) & &-1.7696 (0.0954) & -2.0322 (0.0955) & &  0.4483 (0.1639) &0.4588 (0.1484) & & -1.1991 (0.1275) &-1.7069 (0.1547) & & -5.1424 (0.2583) & -4.7176 (0.2004)   \\
			child$_{\text{m15-m24}_{it}}$& -0.9534 (0.0951) & -0.8537 (0.0911) & & -1.0065 (0.0943) & -0.8928 (0.0905) & & 1.1439 (0.1638) &  1.2409 (0.1407) & & -0.0324 (0.1450) &0.1298 (0.1807) & & -3.9962 (0.1868) & -3.2772 (0.1532)  \\
			child$_{\text{m25-m59}_{it}}$& -0.5194 (0.0752) & -0.3110 (0.0726) & & -0.4647 (0.0745) & -0.3023 (0.0722) & &  0.9204 (0.1353) & 1.3331 (0.1124) & & -0.4053 (0.1144) & -0.4787 (0.1416) & & -2.0476 (0.1501) & -1.5772 (0.1333) \\ \hline \\[-0.5em]
			employ$_{it-1}$& 3.0269 (0.0351) &2.8627 (0.0366)
			& &3.2220 (0.0339)  & 3.0505 (0.0357) & & 2.8941 (0.0658)  &2.6518 (0.0709) & & 2.9531 (0.0598) & 2.7764 (0.0636) & & 3.1070 (0.0554) & 3.0008 (0.0548) \\
			child$_{\text{m1-m14}_{it}}$& -0.7795 (0.0579) & -0.8754 (0.0543) & & -0.7625 (0.0578)  &-0.7935 (0.0534) & & 0.3871 (0.0927)  &0.4901 (0.0842) & &   -0.5653 (0.0933) &-0.6753 (0.0927) & & -2.7104 (0.1471) & -2.6012 (0.1249)    \\
			child$_{\text{m15-m24}_{it}}$& -0.4359 (0.0563) &-0.3529 (0.0532) & & -0.4881 (0.0560)  &-0.3490 (0.0526) & & 0.7339 (0.0984)  & 0.7980 (0.0969) & & -0.3177 (0.1007) &-0.1051 (0.0991) & & -2.0103 (0.1242) & -1.8071 (0.1080)   \\
			child$_{\text{m25-m59}_{it}}$& -0.2282 (0.0450) &-0.1623 (0.0427) & & -0.2172 (0.0448)  &-0.1521 (0.0422) & & 0.4050 (0.0804)  & 0.6896 (0.0763) & & -0.4256 (0.0848) &-0.3117 (0.0810) & & -0.9816 (0.0936) & -0.9927 (0.0987)   \\ \hline \\[-1.0em]
			\multicolumn{15}{l}{\small \textit{Notes}: Robustness checks}	\\
			\multicolumn{15}{l}{\small FE = Fixed Effects, PPL= Penalized Profile Likelihood. All Post Lasso-PPL estimates are bias corrected by half-panel jackknife.} \\
			\multicolumn{15}{l}{\small  Standard errors are provided in (parentheses) and are clustered at the individual level. Estimation without intercept, because all fixed effects are included.}	\\
		\end{tabular}
	\end{center}
\end{sidewaystable*}

\clearpage   
\begin{sidewaystable*}[h!]
	\begin{center}
		\caption{Marginal Effects of Employment Effects}\label{table14}
		\begin{tabular}{p{2.3cm}  >{\raggedleft}p{1.5cm}  >{\raggedleft}p{1.5cm} p{0.1cm}  >{\raggedleft}p{1.5cm}  >{\raggedleft}p{1.5cm} p{0.1cm}   >{\raggedleft}p{1.5cm}   >{\raggedleft}p{1.5cm} p{0.1cm}  >{\raggedleft}p{1.5cm}  >{\raggedleft}p{1.5cm} p{0.2cm}  >{\raggedleft}p{1.5cm} p{1.5cm}<{\raggedleft}}  \hline \\[-1.0em]
			& \multicolumn{2}{c}{LPM FE} & &  \multicolumn{11}{c}{Post Lasso-PPL }  \\[0,05cm] \cline{2-3} \cline{5-15 } \\[-1.0em]
			& \multicolumn{2}{c}{Full sample} & &  \multicolumn{2}{c}{Full sample} & &\multicolumn{2}{c}{Group 1} &  &\multicolumn{2}{c}{Group 2} &  & \multicolumn{2}{c}{Group 3} \\  \cline{2-3} \cline{5-6} \cline{8-9} \cline{11-12} \cline{14-15} \\[-1.0em]
			&	Control mean & Treatment mean &	&	Control mean & Treatment mean &	&	Control mean & Treatment mean & & Control mean & Treatment mean & & Control mean & Treatment mean \\ 
			& (1) & (2) & & (3)  & (4) & & (5) & (6) & & (7) & (8) & & (9) & (10) \\ \hline \\[-1.0em]
			child$_{\text{m1-m14}_{it}}$&  -0.3872 (0.0068)  &-0.4507 (0.0070) & & -0.4270 (0.0090) & -0.4728 (0.0110) & & 0.1182 (0.0044) &0.1065 (0.0047) & &-0.3128 (0.0114) & -0.4658 (0.0171) & & -0.6368 (0.0190) & -0.6694 (0.0181)  \\
			child$_{\text{m15-m24}_{it}}$& -0.2002 (0.0076)  & -0.1918 (0.0078) & & -0.2441 (0.0048) & -0.2131 (0.0042) & & 0.3039 (0.0113) & 0.3006 (0.0105) & &-0.0080    (0.0003) & 0.0336 (0.0013) & & -0.5093 (0.0145) & -0.4869 (0.0122)  \\
			child$_{\text{m25-m59}_{it}}$& -0.1056 (0.0058)  & -0.0681 (0.0060) & & -0.1107 (0.0021) & -0.0712 (0.0013) & & 0.2390 (0.0098) & 0.3039 (0.0157) & & -0.0986 (0.0041) & -0.1205 (0.0043) & & -0.2274 (0.0093) & -0.1960 (0.0070)   \\
			Intercept & 0.5964 (0.0050)  & 0.5742 (0.0051)  & & 0.7186 (0.0146) & 0.6672 (0.0157) & & 0.2344 (0.0300) & 0.1146  (0.0163) & & 0.7496 (0.0229) & 0.7536 (0.0252) & & 0.8964 (0.0185) &  0.8917    (0.0167)   \\ \hline \\[-0.5em]
			employ$_{it-1}$& 0.8399 (0.0031) & 0.8187 (0.0034) & & 0.8254 (0.0044) & 0.8083 (0.0035) & & 0.8021 (0.0048) & 0.7420 (0.0071) & & 0.7876 (0.0096) & 0.7913 (0.0062) & & 0.6347 (0.0094) & 0.6274 (0.0084) \\
			child$_{\text{m1-m14}_{it}}$& -0.0624 (0.0038)  & -0.0817 (0.0043)   & & -0.1016 (0.0010) &-0.1153 (0.0008) & & 0.0561 (0.0009) &0.0758 (0.0014) & & -0.0840 (0.0014) &-0.1106 (0.0008) & & -0.4203 (0.0025) &  -0.4244 (0.0026)  \\
			child$_{\text{m15-m24}_{it}}$&  -0.0321 (0.0041) &-0.0295 (0.0045)  & & -0.0591 (0.0007) & -0.0450 (0.0004) & & 0.1164 (0.0019) & 0.1354 (0.0026) & & -0.0435 (0.0008) & -0.0151 (0.0001) & & -0.2715 (0.0029) & -0.2418 (0.0028)  \\
			child$_{\text{m25-m59}_{it}}$& -0.0162 (0.0031)  & -0.0127 (0.0034)  & & -0.0260 (0.0003) &-0.0194 (0.0002) & & 0.0618 (0.0007) & 0.1182 (0.0019) & & -0.0587 (0.0010) & -0.0447 (0.0004) & & -0.1036 (0.0012) & -0.1069 (0.0013)  \\
			Intercept &0.0989 (0.0032) & 0.1075 (0.0035)  & & 0.1824 (0.0074) & 0.1689 (0.0066)  & & 0.0413 (0.0047) &0.0258  (0.0030) & & 0.2455 (0.0159) & 0.1971 (0.0118) & & 0.5290 (0.0221) & 0.5545    (0.0212)     \\ \hline \\[-1.0em]
		\multicolumn{15}{l}{\small \textit{Notes}: Robustness checks. LPM = Linear Probability Model, FE = Fixed Effects, PPL= Penalized Profile Likelihood. }	\\
	\multicolumn{15}{l}{\small Standard errors are provided in (parentheses), and are clustered at the individual level for the LPM FE (columns (1) and (2)).} \\
	\multicolumn{15}{l}{\small All standard errors and average effects derived from the Post Lasso-PPL estimates are build on a plug-in estimator.}	\\
	\multicolumn{15}{l}{\small See the Appendix for further information on the plug-in principle used here.}	\\
	\multicolumn{15}{l}{\small Intercept is calculated ex-post as the average over all individual fixed effects belonging to the different groups.}	
		\end{tabular}
	\end{center}
\end{sidewaystable*}

\clearpage 
\begin{sidewaystable*}[h!]
	\begin{center}
		\caption{Prediction of Average Employment Probability}\label{table15}
		\begin{tabular}{p{2.3cm} >{\raggedleft}p{1.4cm}  >{\raggedleft}p{1.4cm} >{\raggedleft}p{1.4cm} p{0.1cm}   >{\raggedleft}p{1.4cm}  >{\raggedleft}p{1.4cm} >{\raggedleft}p{1.4cm} p{0.1cm} >{\raggedleft}p{1.4cm} >{\raggedleft}p{1.4cm} >{\raggedleft}p{1.4cm} p{0.1cm} >{\raggedleft}p{1.4cm} >{\raggedleft}p{1.4cm} p{1.4cm}<{\raggedleft}}  \hline \\[-1.0em]
			&  \multicolumn{15}{c}{Employment Probability}  \\[0,05cm]
			& \multicolumn{3}{c}{Full sample} & & \multicolumn{3}{c}{Group 1} &  &\multicolumn{3}{c}{Group 2} &  & \multicolumn{3}{c}{Group 3} \\  \cline{2-4} \cline{6-8} \cline{10-12} \cline{14-16} \\[-1.0em]
			&	Control mean & Treatment mean & $\Delta_{\text{T-C}}$ in p.p. & &	Control mean & Treatment mean & $\Delta_{\text{T-C}}$ in p.p. & & Control mean & Treatment mean & $\Delta_{\text{T-C}}$ in p.p.  & & Control mean & Treatment mean & $\Delta_{\text{T-C}}$ in p.p.  \\ 
			& (1) & (2) & (3) & & (4) & (5) & (6) & & (7) & (8) & (9) & & (10) & (11) & (12) \\ \hline \\[-1.0em]
			child$_{\text{m1-m14}_{it}}$& 0.2689 (0.0145) & 0.1633 (0.0106) &-10.56 [0.0000] & & 0.3409 (0.0352) & 0.1928 (0.0226) &-14.81 [0.0000] & & 0.4411 (0.0283) & 0.2761 (0.0276) & -16.50 [0.0000] & &0.0948 (0.0145) & 0.0843 (0.0106) & -1.05 [0.0244]  \\
			child$_{\text{m15-m24}_{it}}$& 0.4592 (0.0166)& 0.4302 (0.0159) &-2.9 [0.0001] & & 0.5314 (0.0370) & 0.3762 (0.0314) & -15.52  [0.0000] & & 0.7424 (0.0232) & 0.7833 (0.0239) & 4.09 [0.0001] & & 0.2401 (0.0247) & 0.3095 (0.0241) & 6.94 [0.0000]  \\
			child$_{\text{m25-m59}_{it}}$& 0.6041 (0.0162)& 0.5906 (0.0163) &-1.35 [0.0004] & &  0.4685 (0.0374) & 0.4006 (0.0320) & -6.79  [0.0000] & & 0.6529 (0.0262) & 0.6274 (0.0288) & -2.55 [0.0001] & & 0.6223 (0.0280) & 0.6677 (0.0253)  & 4.54 [0.0000] \\
			child$_{\text{m60-m72}_{it}}$& 0.7186 (0.0146)& 0.6672 (0.0157) &-5.14 [0.0000] & & 0.2344 (0.0300) & 0.1146 (0.0163) & -11.98  [0.0000] & & 0.7496 (0.0229) & 0.7536 (0.0252) & 0.40 [0.6735] &  & 0.8964 (0.0185) & 0.8917 (0.0167) & -0.47 [0.4969] \\	\hline \\[-1.0em]
			Ave. emp. & 0.5358 (0.0152) &0.4945 (0.0145) & -4.13 [0.0000] & & 0.4092 (0.0353) &0.3036 (0.0268) & -10.56 [0.0000] & & 0.6388 (0.0253) &0.5990 (0.0264) &-3.98 [0.0000] & & 0.5104 (0.0205) & 0.5387 (0.0188)  & 2.83 [0.0000]                         \\ \hline  \\[-0.5em]
			child$_{\text{m1-m14}_{it}}$& 0.2647 (0.0160) & 0.1605 (0.0104)  & -10.42 [0.0000] & & 0.3225 (0.0330) & 0.1995 (0.0221) &-12.30 [0.0000] & & 0.4217 (0.0323) &0.2116 (0.0223)  & -21.01 [0.0000] & & 0.1469 (0.0196) & 0.1105 (0.0138) & -3.64 [0.0000] \\
			child$_{\text{m15-m24}_{it}}$& 0.4833 (0.0233) & 0.4508 (0.0224) & -3.25 [0.0037] & & 0.5883 (0.0475) &0.4090 (0.0438) & -17.93 [0.0000] & & 0.7213 (0.0365) &0.6752 (0.0400) & -4.61 [0.0188] & & 0.3039 (0.0317) & 0.3616 (0.0311) & 5.77 [0.0001]   \\
			child$_{\text{m25-m59}_{it}}$& 0.6031 (0.0185) & 0.5966 (0.0184) & -0.65 [0.2699] & & 0.5033 (0.0388) &0.4775 (0.0367) & -2.58 [0.0347] & & 0.6075 (0.0318) &0.5742 (0.0307) & -3.33 [0.0029] & & 0.6363 (0.0278)& 0.6550 (0.0274) & 1.87 [0.0212]  \\
			child$_{\text{m60-m72}_{it}}$& 0.7296 (0.0198) & 0.6755 (0.0208) &-5.41 [0.0000] & & 0.2516 (0.0366) &0.1248 (0.0226) & -12.68 [0.0000] & & 0.7992 (0.0287) &0.7307 (0.0323) & -6.85 [0.0000] & & 0.8914 (0.0198) & 0.8888 (0.0197) & -0.26 [0.7616] \\	\hline \\[-1.0em]
			Ave. emp. & 0.5397 (0.0151) &0.5011 (0.0144) & -3.86 [0.0000] &  & 0.4330 (0.0319) &0.3482 (0.0274) &-8.48 [0.0000] & & 0.6190 (0.0263) & 0.5414 (0.0236) & -7.76 [0.0000] & &  0.5356 (0.0213) & 0.5446 (0.0205)  & 0.90 [0.1347]                                                                       \\ \hline  \\[-1.0em]
			\multicolumn{16}{l}{\small \textit{Notes}: Average Employment Probability is calculated as $1/(N[\hat{\mathcal{G}}^k] T[p^j]) \ G(\boldsymbol{x}_{i[\hat{\mathcal{G}}^k]t[p^j]}^{\top} \boldsymbol{\hat{\hat{\alpha}}}_{\hat{\mathcal{G}}^k}+\hat{\mu}_i[\hat{\mathcal{G}}^k] $), where $G(\cdot)$ denotes the CDF of the standard normal,} 	\\
			\multicolumn{16}{l}{\small $p^j, \ j=1,\dots, 4$ denote the four different phases of childage, $[\cdot]$ denotes the restiction to a subspace, and the remainder is defined as above.}	\\
			\multicolumn{16}{l}{\small Standard errors are given in (parentheses) and are build on a plug-in estimator.}  \\
			\multicolumn{16}{l}{\small The $p$-values for the test in eq. \ref{eq:4} are reported under the estimated differences of the average employment predictions.}	
		\end{tabular}
	\end{center}
\end{sidewaystable*}

\clearpage

\section*{Online Appendix}
This online appendix is composed of two main parts. The first part provides further insights into the sample restriction, the descriptives, and the data preprocessing. The second part gives further insights into the underpinning principles of this paper's methodology and estimation.
\subsection*{More on the Data}
\subsubsection*{Restriction of the Sample}
The C-Lasso method of \cite{Su16} does not perform well if $T$ is fixed and $N \rightarrow \infty$, i.e., the number of cross-sectional units is much larger than the number of time periods. In my dataset, there are $57\,453$ women for the considered period. So $N$ is approximately $T^3/7$ ($T=73$), making it hard for the C-Lasso to work well. Furthermore, I do not want to unnecessarily complicate the interpretation of group-specific coefficients by mixing mothers and non-mothers in my analysis. For these reasons, and because I want to identify the reform effect within a regression discontinuity framework, I only focus on mothers with a first child born between October 1, 2006, and March 31, 2007, which drastically reduces my sample size. In total, I consider sample sizes of $N=329$ and $N=333$ for pre- and post-reform samples, respectively.\footnote{The numbers indicate the final sample sizes after preprocessing.} At the same time, I construct a monthly balanced panel to increase the number of time periods and achieve more precise estimates.\footnote{Like \cite{Su16} and \cite{Wange2019}, I generate a balanced panel to simplify the coding.} For the same reasons, I run C-Lasso separately for the pre- and post-reform samples.\\
A next difficulty is that I cannot include mothers who have never worked or who have always worked over the considered period of $T=73$ months (I consider $T=72$ months after first birth) since they are associated with infinite ($-\infty$, $\infty$) fixed effects in a fixed effect probit model. Mothers who have always worked after first birth are missing in my sample because mothers are not allowed to work for two months after birth due to mandatory maternity leave. Mothers who have never worked over the considered period are a bigger issue. My full sample (pre- and post-reform samples) contains 123 (out of 785) mothers who have never worked over the 72 months after the first birth, which is 15,7\%. Conversely, this means that I do not look at mothers who completely concentrate on the family work and are never employed after birth. The mothers I look for have a certain attachment to the labor market and return to employment at some point, i.e., I examine the heterogeneous dynamics of when women return to work after birth. \\
\textbf{Descriptives} Table \ref{table11} shows the average employment rate in \% and the percentage of mothers never employed for the respective phases up to six years after childbirth. A graphical illustration of the average employment rate can be found in Section \ref{Data}. It can be already seen descriptively that a larger share of mothers is employed in the medium-run (child$_{\text{m25-m59}_{it}}$) post-reform. Indeed, \cite{Kluv18} have found that the parental benefit reform caused a relatively large share of mothers to move from nonparticipation to employment in the medium run. Moreover, the descriptive statistics show that a lower percentage of women are `never employed' during the period under consideration.
\begin{table*}[h!]
	\begin{center}
		\caption{Descriptives}\label{table11}
		\begin{tabular}{p{2.4cm}   >{\raggedleft}p{2.2cm}  >{\raggedleft}p{2.3cm} p{0.2cm} >{\raggedleft}p{2.2cm} p{2.3cm}<{\raggedleft}}  \hline \\[-1.0em]
			&\multicolumn{2}{c}{Pre-Reform Sample} & &\multicolumn{2}{c}{Post-Reform Sample}  \\  \cline{2-3} \cline{5-6} \\[-1.0em]
			&Average employment rate in \% & $\%$ of women never employed & &Average employment rate in \% & $\%$ of women never employed	 \\ \hline \\[-1.0em]
			child$_{\text{m1-m14}_{it}}$ & 20.9 & 56.5 & & 12.4 & 54.5  \\[0,05cm]
			child$_{\text{m15-m24}_{it}}$ & 39.6 & 50.8 & & 38.2 & 49.6 \\[0,05cm]
			child$_{\text{m25-m59}_{it}}$ & 49.1 & 25.4 & & 50.6 & 22.0 \\ [0,05cm]
			child$_{\text{m60-m72}_{it}}$ & 59.6 & 30.0 & & 57.4 & 30.0 \\ [0,05cm]
			$\Sigma$          & 43.9 & 17.3 & & 42.3 & 14.0 \\[0,05cm] \hline \\[-1.0em] 
			\multicolumn{6}{l}{\small \textit{Notes:} Descriptives based on data without the preprocessing step.}			
		\end{tabular}
	\end{center}
\end{table*}

\subsubsection*{Covariates}
\textbf{Stationarity}
A major concern in the model is the underlying assumption of stationarity. One may cast doubt that the covariates are stationary. Obviously, the cross-sectional distribution of the child age dummies of the first child changes over time, but also the age of the second child is clearly trending upwards over the sampling period. Controlling for the second child's age shows up in the form of diverging estimates across the two subpanels and leads to a huge bias estimated by the split-panel jackknife. To reduce the second child age variable's strong upward trend, I have decided to control for a second child with a dummy variable. \\
Another problem is the assumption that the initial observations on employment are drawn from a steady-state distribution (In my data preparation, all women give birth to a first child at time $t=0$, that is, no women are employed at $t=0$.). Therefore it is unlikely that the initial observations on employment are drawn from a steady-state distribution.  \\ 
\textbf{Exogeneity} One advantage of my data preparation is that I do not have to worry about the exogeneity assumption of the first child. Nevertheless, I assume exogeneity for the second child. However, in his seminal paper on the interaction between labor-market and fertility decisions, \cite{Hys99} is unable to reject the exogeneity of fertility decisions once lagged employment decisions are taken into account. This finding suggests that the exogeneity assumption should not be a cause for major concern in my model.

\subsection*{More on the Methodology}
\subsubsection*{General}
All fixed effects are included in the FE estimation, and therefore the constant is left out.
\subsubsection*{Group comparisons}
Since I run C-Lasso separately for the pre- and post-reform samples, I have to make assumptions and formulate criteria to compare the different groups appropriately. Applying C-Lasso on the dataset, I find three latent groups for both samples. This has the advantage that I can find an associated control group for each treatment group. I apply a distance-based method to find the appropriate treatment and control groups for comparison.\footnote{I choose the Minkowski-distance as a suitable measure $d(\boldsymbol{x},\boldsymbol{x}')=(\sum_{i=1}^n |x_i - x_i'|^p)^{1/p} = ||\boldsymbol{x}-\boldsymbol{x}'||_p$ and set $p = 1$.} The aim is to compare those control and treatment groups that are most similar in terms of the number of individuals classified into the different groups and the predicted average employment probabilities. Thus, the pair $(\hat{\mathcal{G}}^{k^T},\hat{\mathcal{G}}^{k^C})$ of treatment and control groups to be compared obey the following minimization rule:
\begin{align*}
A &:= \min_{o_1,\dots,o_{3!}  }\{||\boldsymbol{N}^T_{o_j[\hat{\mathcal{G}}^k]}- \boldsymbol{N}^C_{o_j[\hat{\mathcal{G}}^k]}||_1\} \ \land \ B := \min_{o_1,\dots,o_{3!}  }\{||\overline{\mbox{$\boldsymbol{employ}$\raisebox{2mm}{}}}^T_{o_j[\hat{\mathcal{G}}^k]}- \overline{\mbox{$\boldsymbol{employ}$\raisebox{2mm}{}}}^{ C}_{o_j[\hat{\mathcal{G}}^k]}||_1\}, \\  k&=1,\dots,3, j = 1, \dots, 3!,
\end{align*}
where $o_1,\dots,o_{3!}$ are permutations on $\{\hat{\mathcal{G}}^1_T, \hat{\mathcal{G}}^2_T, \hat{\mathcal{G}}^3_T, \hat{\mathcal{G}}^1_C, \hat{\mathcal{G}}^2_C, \hat{\mathcal{G}}^3_C\} $, $\boldsymbol{N}$ is a vector containing the number of mothers in each group, $\overline{\mbox{$\boldsymbol{employ}$\raisebox{2mm}{}}}$ is a vector containing the predicted average employment probabilities over the different phases of child age, and the rest of the notation is defined as above. Table \ref{table12} shows that the first permutation $o_1$ is the optimal choice for finding the treatment and control group pairs. This permutation also corresponds to the order of the group classification of the C-Lasso, i.e. we receive the pairs $(\hat{\mathcal{G}}^{1^T},\hat{\mathcal{G}}^{1^C}),(\hat{\mathcal{G}}^{2^T},\hat{\mathcal{G}}^{2^C}),(\hat{\mathcal{G}}^{3^T},\hat{\mathcal{G}}^{3^C})$.
\begin{table*}[h!]
	\begin{center}
		\caption{}\label{table12}
		\begin{tabular}{p{1cm}   >{\raggedleft}p{2cm} p{2cm}<{\raggedleft}}  \hline \\[-1.0em]
			& A  &  B \\ \hline \\[-1.0em]
			$o_1$  & 28 & 0.8626   \\[0,05cm]
			$o_2$ & 58 & 2.1606 \\[0,05cm]
			$o_3$ & 230 & 2.5724 \\ [0,05cm]
			$o_4$ & 176 & 1.7588 \\ [0,05cm]
			$o_5$ & 230 & 3.0338                           \\ [0,05cm]
			$o_6$ & 206 & 2.5954                      \\[0,05cm] \hline \\[-1.0em] 				
		\end{tabular}
	\end{center}
\end{table*}   

\subsubsection*{Split-Panel Jackknife}
In this subsection, I introduce the basic idea of the half-panel jackknife (\cite{Dhaen15}) method to reduce the bias in fixed-effect models. It is well known that, in particular, maximum-likelihood estimates of nonlinear dynamic models with fixed effects are subject to the incidental parameter problem (\cite{Ney48}). I follow \cite{Su16} and use the half-panel jackknife to correct the bias in the post-C-Lasso estimates. The jackknife-correction method is explicitly designed to handle dynamics in the data and yields estimators that are simple to implement, requiring only a few maximum-likelihood estimates.\\
In my empirical application, I consider $T=73$ and partition the panel into two half-panels. As $T$ is odd, I split the panel into non-overlapping half-panels of the form
\begin{align*}
\mathcal{S} = \{S_1,S_2\}, \quad \text{where} \quad S_1:=\{1,2,\dots, \floor*{73/2}   \}, \ S_2:=\{\floor*{73/2}+1, \dots, 73\}.
\end{align*}
The half-panel estimator $\tilde{\theta}_{1/2}$ proposed by \cite{Dhaen15} is then defined as follows
\begin{align*}
\tilde{\theta}_{1/2} := 2 \hat{\theta} - \bar{\theta}_{1/2}, \quad \bar{\theta}_{1/2}:=\frac{1}{2}(\bar{\theta}_{\mathcal{S}_1}+\bar{\theta}_{\mathcal{S}_2}),
\end{align*}
with $\bar{\theta}_{\mathcal{S}_1}:=\sum_{S_1 \in \mathcal{S}_1} \frac{|S_1|}{36} \hat{\theta}_{S_1} $ and $\bar{\theta}_{\mathcal{S}_2}:=\sum_{S_2 \in \mathcal{S}_2} \frac{|S_2|}{37} \hat{\theta}_{S_2} $. 
\subsubsection*{Plug-in Estimation}
In this subsection, I introduce the plug-in estimators for the mean and variance. Plug-in estimation is a general scheme to estimate parameters. I follow \cite{Dhaen15} and use this estimation scheme for computing average marginal effects (AME) and its standard error estimates, the parameters of main interest in my empirical approach. An introduction of the plug-in estimation can be found, for instance, in \cite{Lederer2018}.\footnote{cf. Chapter 3 on Pages 43ff}\\
First, I consider the population mean 
\begin{align*}
\gamma := \mathbb{E}x
\end{align*}
of a random variable $x:(\mathcal{A}, \mathfrak{A}, \mathbb{P})  \rightarrow (\mathbb{R}, \mathfrak{B})$. Setting $f[\mu]:= \int \text{id} \ d\mu, \ \text{id}[a]=a$ the identity function, for every probability measure $\mu$ on $\mathfrak{B}$, we find 
\begin{align*}
\gamma = f[x[\mathbb{P}]],
\end{align*}
which brings us into the plug-in framework. Given data $x^1,\dots,x^n$, estimating $x[\mathbb{P}]$ by the emprirical measure yields the empirical mean
\begin{align*}
\gamma = f[\hat{\mathbb{P}}_n]= \int  \text{id} \ d\hat{\mathbb{P}}_n = \frac{1}{n} \sum_{i=1}^n x^i.
\end{align*}
I look on fixed-effect plug-in estimates of the AME for continuous and discrete variables. Assume that $x_{1_{it}}$  is a binary variable and $x_{2_{it}}$ is a continuous variable. The AME for $x_{1_{it}}$ is obtained by
\begin{align}\label{eq:25}
AME_{x_{1_{it}}} = &\frac{1}{NT} \sum_{i=1}^N \sum_{t=1}^T [G(\hat{\beta}_1+ x_{2_{it}}\hat{\beta}_2+ \dots +x_{5_{it}}\hat{\beta}_5 +\hat{\mu}_i)-G(x_{2_{it}}\hat{\beta}_2+ \dots +x_{5_{it}}\hat{\beta}_5 +\hat{\mu}_i)],
\end{align}
where $G(\cdot)$ denotes the standard normal CDF.
The AME for the continuous variable $x_{2_{it}}$ is obtained by
\begin{align}\label{eq:26}
AME_{x_{2_{it}}} = &\frac{1}{NT} \sum_{i=1}^N \sum_{t=1}^T g(\hat{\beta}_1+ x_{2_{it}}\hat{\beta}_2+ \dots +x_{5_{it}}\hat{\beta}_5 +\hat{\mu}_i) \hat{\beta}_2 ,
\end{align}
where $g(\cdot)$ denotes the standard normal probability density function (PDF). Thus, for computing AME, I simply use the plug-in estimate of the population mean. \\ 
Second, we consider the population variance
\begin{align*}
\sigma^2 := \mathbb{E}[(x-\mathbb{E}x)^2]
\end{align*}
of a random variable $x:(\mathcal{A}, \mathfrak{A}, \mathbb{P})  \rightarrow (\mathbb{R}, \mathfrak{B})$. Setting $f[\mu]:= \int \text{id} \ d\mu, \ \text{id}[a]=a$ the identity function, for every probability measure $\mu$ on $\mathfrak{B}$, we find 
\begin{align*}
\sigma^2 = f[x[\mathbb{P}]],
\end{align*}
which brings us into the plug-in framework. Given data $x^1,\dots,x^n$, estimating $x[\mathbb{P}]$ by the emprirical measure yields the empirical variance
\begin{align*}
\sigma^2 = f[\hat{\mathbb{P}}_n]= \int  \text{id} \ d\hat{\mathbb{P}}_n = \frac{1}{n} \sum_{i=1}^n (x^i-\bar{x})^2,
\end{align*}
with $\bar{x}=\sum_{i=1}^n x^i$. \\
Because the population mean is not known in practice, I replace the population variance with the sample variance. The standard error estimates are finally obtained by taking the square root of the sample variance. Thus, I look at plug-in estimates of the cross-sectional variance of the within-group AME:
\begin{align}
s^2 := \frac{1}{N-1} \sum_{i=1}^N (AME_i -AME)^2, \quad \text{with} \quad  AME=\frac{1}{N} \sum_{i=1}^N  AME_i. 
\end{align} 
\cite{Dhaen15} notice that the estimators in \ref{eq:25} and \ref{eq:26} are subject to two sources of bias, even if a bias-corrected estimator (here I use the split-panel jackknife estimator) is used instead of the maximum-likelihood estimator. The first stems from using $\hat{\mu}_i$ instead of $\mu_i$, and the second arises from using ($\boldsymbol{\hat{\gamma}}, \boldsymbol{\hat{\beta}}_i$) instead of ($\boldsymbol{\gamma}, \boldsymbol{\beta}_i$). Combined with the bias in the AME, this results in confidence intervals with poor coverage. However, the authors show both analytically and with simulations that this is a problem when $T$ is relatively small compared to $N$. Under rectangular-array asymptotics, when $N$ is fixed and $T \rightarrow \infty$, the correct inference is obtained even without bias correction. Since I consider a relatively large $T$ compared to my sample size $N$ in my empirical application, the bias is almost negligible, and I refer to the asymptotic equivalence of the estimated AME with and without bias-correction. Therefore, I do not perform some bias correction when estimating the AME and its standard errors.


\end{document}